\documentclass[a4paper,onecolumn,10pt,accepted=2025-10-05]{quantumarticle}
\pdfoutput=1
\usepackage[utf8]{inputenc}
\usepackage[english]{babel}
\usepackage[T1]{fontenc}
\usepackage[usenames,dvipsnames]{xcolor}
\usepackage{amssymb}
\usepackage{graphicx}
\usepackage{amsmath}
\usepackage{dsfont}
\usepackage{mathtools} 
\mathtoolsset{showonlyrefs} 
\usepackage{amsthm}
\usepackage[bookmarks=true,colorlinks,citecolor=blue,urlcolor=blue]{hyperref}
\newtheorem{thm}{Theorem}[section]      

\usepackage{braket}
\usepackage{blindtext}
\usepackage{physics}
\usepackage{float}
\usepackage{enumitem}
\usepackage[numbers,sort&compress]{natbib}
\usepackage{bm}

\newcommand{\Bl}{\Bigg( \Bigg.}
\newcommand{\Br}{\Bigg) \Bigg.}
\newcommand{\bl}{\bigg( \bigg.}
\newcommand{\br}{\bigg) \bigg.}
\newcommand{\Bsl}{\Bigg[ \Bigg.}
\newcommand{\Bsr}{\Bigg] \Bigg.}

\begin{document}

\author{Elisabeth Wybo}
\email{elisabeth.wybo@meetiqm.com}
\affiliation{IQM Quantum Computers, Georg-Brauchle-Ring 23-25, 80992 München, Germany}
\author{Martin Leib}
\email{martin.leib@meetiqm.com}
\affiliation{IQM Quantum Computers, Georg-Brauchle-Ring 23-25, 80992 München, Germany}

\title{Missing Puzzle Pieces in the Performance Landscape of the Quantum Approximate Optimization Algorithm}

\begin{abstract}
We consider the maximum cut and maximum independent set problems on random regular graphs in the infinite-size limit, and calculate the energy densities achieved by QAOA for high degrees up to $d=100$. Such an analysis is possible because the reverse causal cones of the operators in the Hamiltonian are with high probability associated with tree subgraphs, for which efficient classical contraction schemes can be developed. We combine the QAOA analysis with state-of-the-art upper bounds on optimality for both problems. This yields novel and better bounds on the approximation ratios achieved by QAOA for large problem sizes. We show that the approximation ratios achieved by QAOA improve as the graph degree increases for the maximum cut problem. However, QAOA exhibits the opposite behavior for the maximum independent set problem, i.e. the achieved approximation ratios decrease when the degree of the problem is increased. This phenomenon is explainable by the overlap gap property for large $d$, which restricts local algorithms (like QAOA) from reaching near-optimal solutions with high probability. In addition, we use the QAOA parameters determined on the tree subgraphs for small graph instances, and in that way outperform classical algorithms like Goemans-Williamson for the maximum cut problem and minimal greedy for the maximum independent set problem. In this way we circumvent the parameter optimization problem and are able to compute the expected approximation ratios.
\end{abstract}

\maketitle

\section{Introduction}

The advent of quantum computing promises advantage over classical computing in areas as diverse as quantum simulation~\cite{Feynman1982,Cirac2012,Georgescu2014}, machine learning~\cite{Harrow2009,Liu2021,Huang2021,Sweke2021} and optimization~\cite{Grover1996,Abbas2023}. To unequivocally profit from this however, fault tolerant operation of quantum computers is needed. In the current era of Noisy Intermediate Scale Quantum (NISQ) devices~\cite{Preskill2018}, it is not possible to run the deep quantum algorithms equipped with proofs of quantum advantage. Therefore, a new research topic of quantum heuristics with substantially reduced number of required gate operations is currently flourishing. The reduced number of gates for these heuristics comes typically at the cost of unclear prospects for quantum advantage caused by a lack of techniques for their complexity analysis. Compared to other quantum algorithms for the NISQ era, there is however already a lot known about performance and time-to-solution for the Quantum Approximate Optimization Algorithm (QAOA)~\cite{Farhi2014,Farhi2014a,Farhi2016,Hadfield2017,Zahedinejad2017,Wang2018,Blekos2023} which the current work is adding a missing puzzle piece to.

Two basic insights have enabled analytical investigations of the average performance of the QAOA: First, the fact that the expected performance of QAOA is the sum over terms which are only affected by a subset of gates, called the Reverse Causal Cone (RCC), which is limited by the number of layers in the QAOA circuit and not the actual problem size. Second, the focus on either all-to-all or random regular problem graphs. 
Random $d$-regular graphs have the convenient property that the $p$-local environments of the nodes and edges, are asymptotically (with growing number of graph nodes $N$) trees. This is because the probability of having small loops vanishes~\cite{Makover2006}. Consequentially the topology of the qubits involved in a single RCC of QAOA is also, with high probability a tree. Therefore, it follows that the asymptotic performance of QAOA with depth $p$, is fully determined by its performance on those tree structures~\cite{Farhi2014,Basso2021}, which allow to use efficient classical contraction schemes. These locality arguments have already been explored in the original QAOA paper for $3$-regular graphs~\cite{Farhi2014}. Additionally, it was also possible to calculate optimal, instance- and problem size-independent, parameters thereby creating an optimization loop free variation of QAOA termed `tree QAOA'~\cite{Brandao2018,Streif2020,Wurtz2020,Galda2021,Galda2023}. 


With the help of the above described techniques, it was possible to show that QAOA outperforms on average the classical worst-case performance guarantee of the Goemans-Williamson (GW) algorithm for the Maximum Cut (MaxCut) problem on high-girth regular graphs, if $p$ is large enough~\cite{Crooks2018,Wurtz2021}. The GW algorithm~\cite{Goemans1995} is the optimal efficient classical approximate algorithm for this problem with a performance guarantee, assuming that the unique games conjecture is true~\cite{Khot2007}.
It should be noted, however, that some classical algorithms, while lacking a worst-case performance guarantee, perform better on average than the GW guarantee~\cite{Hastings2019,Guerreschi2019,Lykov2022,Alaoui2021}. In particular, in Ref.~\cite{Alaoui2021} it was shown that an approximate message passing algorithm finds solutions that are arbitrarily close to the optimal solution in the large degree limit. Also in Ref.~\cite{Barak2021} it was shown that there exists a classical polynomial time algorithm that can beat QAOA in many cases. For the MaxCut problem, the asymptotic QAOA angles have been determined explicitly for the smallest $p$~\cite{Farhi2014,Wang2018,Wurtz2020,Marwaha2021a}, either using numerical contraction~\cite{Streif2020, Wurtz2021}, or by evaluating recursive formulas that describe the contraction sequence analytically~\cite{Farhi2022,Basso2021,Boulebnane2021}. The numerical approach of Refs.~\cite{Streif2020, Wurtz2021} is limited to fairly small degrees. On the other hand, in the approach of Ref.~\cite{Basso2021} the degree does not enter in the complexity of the evaluation of the variational energy. The tree angles have been evaluated up to $p = 11$ for MaxCut on 3-regular graphs~\cite{Wurtz2021}. This also implies that the asymptotic performance of QAOA on this problem is known explicitly up to depth $p=11$. However, the performance scaling with $p$ is not formally known. Summarizing, it can be said that fixed depth QAOA exhibits a constant, non-vanishing performance as function of the problem size $N$, which can also be extended to Sherrington-Kirkpatrick models on all-to-all connected problem graphs~\cite{Farhi2022}. 

In contrast to this relative success of QAOA in solving the MaxCut problem, it has been explicitly shown that constant depth QAOA has strong sub-optimal performance for the Maximum Independent Set (MIS) problem~\cite{Farhi2020a,Farhi2020}. Indeed, the minimal requirement to find solutions close to the optimum is `to see the whole graph', thus that $p\in O(\log(N))$, making QAOA an algorithm with increasing query complexity. This observation of widely differing algorithm performance for two NP-hard problem classes, MaxCut and MIS, is not restricted only to quantum algorithms or QAOA but also extents to the performance of classical algorithms. 

A recent, very successful attempt at explaining this phenomenon employs the so-called Overlap Gap Property (OGP) that can be linked to algorithmic hardness. Problems that do have an OGP, like MIS in limit of large graph degree, possess an intricate property associated to the clustering of good solutions~\cite{Mezard2005,Achlioptas2011,Gamarnik2019,Gamarnik2021}. Solutions that are close to optimality for instances which possess OGP are either very similar or very different, implying that there are remote clusters of good solutions in the state space.   

It can be shown that the existence of an OGP implies that there is an a-priori gap between what $p$-local algorithms (like QAOA) can possibly achieve, and optimality~\cite{Chen2023}. In contrast to this, for problems without OGP~\cite{Chen2017,Auffinger2017,Chou2021}, we may be able to find a $p$-local algorithm that produces solutions $(1-\epsilon)$ close to optimality with high probability. The statement `with high probability' is important, and implies that there is no guarantee that such a good $(1-\epsilon)$ approximation can be always reached, as otherwise complexity theoretic statements like the unique games conjecture~\cite{Khot2007}, and therefore also \texttt{P} $\neq$ \texttt{NP}, would collapse. For MaxCut on $d$-regular graphs, such an approximate $(1-\epsilon)$ classical algorithm has been proposed in Ref.~\cite{Alaoui2021}. However, in this algorithm $\epsilon$ cannot be chosen arbitrarily small for finite $d$. The performance gap of QAOA in solving problems with and without OGP becomes bigger for random $d$-regular graphs as $d$ increases. The regime of finite $d\geq 3$ has been treated with QAOA with fixed $p$ for MaxCut with less stringent bounds on optimality~\cite{Wurtz2021}. Such an analysis suggests that the approximation ratios achieved by QAOA in solving MaxCut are decreasing with growing $d$, which is a  qualitatively wrong result. In this work, we show they are not decreasing by using tighter bounds on optimality. For the MIS problem the same regime $d\geq 3$ has not been explicitly investigated yet with QAOA to the best of our knowledge. 

In this work we therefore provide an explicit performance analysis of QAOA on these two problems. For this we extend the technique of the recursive formula of Ref.~\cite{Basso2021} to MIS, enabling performance calculations in the large but finite $d$ regime. We start by showing how to treat the MaxCut problem and the MIS problem on equal footing using the language of Ising models with local fields. With a range of numerical improvements in evaluating the recursive formula, we are able to determine tree QAOA performance for Ising models with local fields $H=\sum_{ij \in E}Z_iZ_j + h\sum_{i\in V}$ for $N\rightarrow \infty$ on $d$-regular graphs with $d$ up to $100$. 

In addition, we combine these results with state-of-art upper bounds on optimality. For MaxCut, upper bounds on the cut fraction in the asymptotic limit for $d$-regular graphs have been determined in Refs.~\cite{Dembo2015,CojaOghlan2020}. For MIS, upper bounds on the independence ratio in the asymptotic limit for $d$-regular graphs have been determined in Refs.~\cite{McKay1987,Balogh2017}. Taking into account these findings yields higher performance guarantees for QAOA in the asymptotic limit than previously assumed, and changes insights on the qualitative behaviour in the large $d$ regime. This is mainly relevant for MaxCut: for instance in Ref.~\cite{Wurtz2021} it is observed that asymptotic tree QAOA outperforms the GW guarantee at $p\geq 11$. However, by combining the QAOA performance with state-of-the-art bounds on optimality, it can be realized that this actually already happens when $p\geq 4$ for $N\rightarrow \infty$. This observation is only true in the large $N$ limit, but however gives hope that an effective quantum utility regime can be reached in the near term for this problem, with a local quantum algorithm.

Combining the better bounds with the extended method of the recursive formula, we are in the end able to show a manifestly different behavior of QAOA in solving the MaxCut and MIS problems for increasing graph degrees. The MIS problem has an OGP regime in the large $d$ limit~\cite{Gamarnik2013}, but such regime is not proven for finite $d$. However, for this problem we observe a decreasing QAOA performance with increasing $d$. On the other hand for MaxCut, we do not observe such a decreasing performance. Despite these findings, there is still hope for MIS that with a quantum algorithm breaking locality, i.e. for QAOA with $p \in \Omega(\log(N))$, near-optimal performances can still be achieved in polynomial-time on a quantum computer. On the other hand, the performance of QAOA with $p \in O(\log(N))$ appears limited as classical algorithms are provably optimal up to the OGP bound~\cite{Chen2023}. However, there is also future hope that hybrid approaches can provide near-optimal performances in overall polynomial time~\cite{Bravyi2019,Bae2022,Patel2022,Wagner2023,Dupont2023a,Dupont2023,Finzgar2023}.


Our paper is organized as follows: In Sec.~\ref{sec:QAOA}, we introduce the QAOA algorithm and discuss its $N\rightarrow \infty$ limit for regular graphs which is called `tree QAOA'. In Sec.~\ref{sec:model}, we discuss the MaxCut and MIS problems and show how they can be unified as Ising models. Here, we also define the performance metrics that we will measure on the QAOA ansatz state in order to get the expected solution quality of the sampled solutions. In Sec.~\ref{sec:OGP}, we briefly review the concept of Overlap Gap Property and local algorithms, and provide a short overview about the different hardness regimes of the two problems, and the implications for QAOA. In Sec.~\ref{sec:results}, we present and discuss our results. The main result of our paper is shown in Fig.~\ref{fig:fixed_h_performance}, where the asymptotic performances of QAOA for both problems are summarized for many different graph degrees. We conclude in Sec.~\ref{sec:concl}, and provide the full tree calculations in Appendix~\ref{sec:tree_calc}.

\section{QAOA and tree QAOA} \label{sec:QAOA}

\begin{figure*} 
\centering
\includegraphics[width=0.95\textwidth]{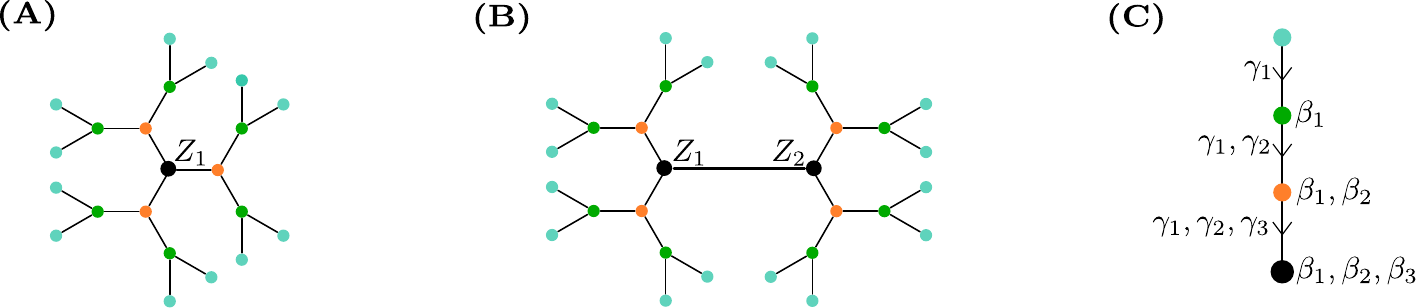}
\caption{(A,B) The 1-tree and 2-tree subgraphs at level $p=3$, for degree $d=3$. (C) The path from outer leaf to root, together with the angle dependencies of each layer. The root variable is shown in black, the other variables are color coded according to their layer: $m=1$ orange, $m=2$ green, $m(=p)=3$ blue.} \label{fig:trees}
\end{figure*}

In this section, we review the QAOA algorithm, and summarize its asymptotic limit in the number of nodes $N$ for $d-$regular graphs, which we will refer to as `tree QAOA'. QAOA is a variational algorithm that is designed to find low-energy states of a diagonal Hamiltonian $H$. Our work considers the Ising Hamiltonian on random $d$-regular graphs $G(V,E)\in \mathbb{G}(N,d) $, with $V$ the vertex set $|V|=N$, and $E$ the edge set $|E|=Nd/2$,
\begin{equation}\label{eq:H_Ising}
H = \frac{1}{\sqrt{d}} \left( \sum_{ij\in E} Z_i Z_j +  h\sum_{i\in V} Z_i \right).
\end{equation}
Here $Z_i$ is the Pauli-$Z$ operator associated with the $i$-th qubit, we use the convention that $Z_i \ket{\bm{z}} = z_i \ket{\bm{z}}$ with $z_i \in \{-1,1\}$ and $\bm{z} = (z_1,\dots,z_N)$. To find low-energy states, one optimizes a depth-$p$ parameterized circuit ansatz that has been introduced in Ref.~\cite{Farhi2014} and that consists of alternating layers of mixing and phase-separator unitaries
\begin{equation} \label{eq:qaoa_ansatz}
\ket{\bm{\gamma},\bm{\beta}} =  e^{-i \beta_p B} e^{-i\gamma_p H}  \dots e^{-i \beta_1 B} e^{-i\gamma_1 H} \ket{+}\,,
\end{equation}
where $B = \sum_i X_i$ is the mixing operator with $X_i$ the Pauli-$X$ operator associated to the $i$-th qubit. The initial state is an eigenstate of the mixing operator, and is defined as $\ket{+} = \frac{1}{\sqrt{2^N}} \sum_{\bm{z}} \ket{\bm{z}}$. Using the Ritz' variational principle one can find optimal parameters $\bm{\beta}=(\beta_1,\dots,\beta_p)$ and $\bm{\gamma}=(\gamma_1,\dots,\gamma_p)$ in a classical outer optimization loop such that the variational energy of the classical cost function, in our case Eq.~\eqref{eq:H_Ising}, is minimized
\begin{equation} \label{eq:qaoa_var}
    \min_{\bm{\gamma},\bm{\beta}} \ev{H}{\bm{\gamma},\bm{\beta}}.
\end{equation}
This guarantees that the measurement samples $\bm{z}$ from the prepared quantum state $\ket{\bm{\gamma},\bm{\beta}}$ on average have low energy $H(\bm{z})=\ev{H}{\bm{z}}$ in a noiseless scenario. Ideally, the optimal solution $\bm{z}_{opt}$, for which $H(\bm{z}_{opt}) = \min_{\bm{z}} H(\bm{z})$, is sampled with non-vanishing probability. The QAOA ansatz resembles a trotterization of the quantum annealing protocol and therefore the adiabatic theorem guarantees that this is true for $p \rightarrow \infty$ with unit probability, but this is not practically attainable. It can however be shown that the variational method, where the trotter-step sizes are optimized, always performs better compared to continuous evolution if the total evolution time is finite~\cite{Yang2017}. Therefore, in this paper we want to investigate which energies~\eqref{eq:qaoa_var} can be reached in the limit of large instances $N\rightarrow \infty$ for \textit{fixed} $p$.

The energy expectation value~\eqref{eq:qaoa_var} consists of a sum of local expectation values, because the summations in Eq.~\eqref{eq:H_Ising} run respectively over the edges and vertices of the problem graph. As the problem graph also determines the two-qubit gates in the QAOA ansatz, it follows that $\ev{Z_i}{\bm{\gamma},\bm{\beta}}$ only depends on gates acting on qubits that are within distance $p$ in the problem graph from vertex $i$. All other gates commute with the measurement operator $Z_i$ and undergo a unitary cancellation. This is true in a similar fashion for $\ev{Z_i Z_j}{\bm{\gamma},\bm{\beta}}$ with $ij \in E$, i.e. only gates acting on qubits that are within distance $p$ from either $i$ or $j$ in the problem graph contribute. 
The qubits that are part of the causal cone of a local operator form a subgraph of the original problem graph. We will refer to the set of gates acting on such a subgraph, and determining the expectation value of a local operator, as the Reverse Causal Cone (RCC) of the operator.
When uniformly sampling large random regular graphs, the probability of having small loops (e.g. triangles) in the graph is exponentially suppressed~\cite{Makover2006}. From this it follows that all the causal-cone subgraphs have the same topology of regular trees when $N\rightarrow \infty$. Therefore, the energy density measured on the QAOA ansatz state reduces to
\begin{equation}\label{eq:energy_density}
\lim_{N\rightarrow \infty} \frac{\ev{H}{\bm{\gamma},\bm{\beta}}}{N} = \frac{\sqrt{d}}{2} \ev{Z_1 Z_2}{\bm{\gamma},\bm{\beta}}^{2\textrm{-tree}} + \frac{h}{\sqrt{d}} \ev{Z_1}{\bm{\gamma},\bm{\beta}}^{1\textrm{-tree}},
\end{equation}
where $H$ is given by Eq.~\eqref{eq:H_Ising}. The expectation values $\ev{Z_1 Z_2}{\bm{\gamma},\bm{\beta}}^{2\textrm{-tree}}$ and $\ev{Z_1}{\bm{\gamma},\bm{\beta}}^{1\textrm{-tree}}$ are defined on a QAOA state which consists of a reduced amount of spin variables that are arranged in tree subgraphs. The edge correlator $\ev{Z_1 Z_2}{\bm{\gamma},\bm{\beta}}^{2\textrm{-tree}}$ describes the expectation value corresponding to a causal-cone subgraph that is a $d$-regular tree of depth $p$ with two roots. Similarly the onsite expectation value $\ev{Z_1}{\bm{\gamma},\bm{\beta}}^{1\textrm{-tree}}$ corresponds to a single-root tree subgraph. For $d=3$ and $p=3$ these causal-cone subgraphs are shown in Fig.~\ref{fig:trees}.

\subsection{Summary of the recursive iterations}
In Appendix~\ref{sec:tree_calc}, we provide recursive formulas to compute the two terms in Eq.~\eqref{eq:energy_density}. These generalize the formulas derived in Ref.~\cite{Basso2021} by breaking the $\mathbb{Z}_2$ symmetry of MaxCut, and by including a block symmetry in the iterations. In this section, we will only summarize the results of~\ref{ssec:add_h}, and refer to the Appendix~\ref{sec:tree_calc} for the full derivation and for more compact iterations in which the number of terms is reduced.

The two terms needed to evaluate Eq.~\eqref{eq:energy_density} can be worked out as
\begin{multline} \label{eq:ZZ_rec}
\ev{Z_1 Z_2}{\bm{\gamma},\bm{\beta}}^{2\textrm{-tree}} \\ = \sum_{\bm{a},\bm{b}}  a_0 b_0 f(\bm{a}) f(\bm{b}) H_{d-1}^{(p)}(\bm{a})  H_{d-1}^{(p)}(\bm{b})  \exp(i \frac{1}{\sqrt{d}}  \bm{\Gamma} \cdot  \bm{a}\bm{b}) \exp(i \frac{h}{\sqrt{d}} \bm{\Gamma} \cdot  (\bm{a}+\bm{b})),
\end{multline}
and 
\begin{equation}  \label{eq:Z_rec}
\ev{Z_1}{\bm{\gamma},\bm{\beta}}^{1\textrm{-tree}} = \sum_{\bm{a}}  a_0 f(\bm{a}) H_{d}^{(p)}(\bm{a})   \exp(i \frac{h}{\sqrt{d}} \bm{\Gamma} \cdot \bm{a}).
\end{equation}
Here the summation $\sum_{\bm{a}}$ ($\sum_{\bm{a},\bm{b}}$) means summing over one basis set (two independent basis sets) and runs over all possible vectors 
\begin{equation}
\bm{a} = ( a_1, \dots a_p,a_0,a_{-p},\dots a_{-1}),\; a_{i} \in \{-1,1\},
\end{equation}
describing bitstrings of length $2p+1$ (similar for $\bm{b}$). Following the notation of Ref.~\cite{Basso2021}, we also arrange the QAOA angles corresponding to the phase separator in a vector of length $2p+1$
\begin{equation}
\bm{\Gamma} = (\gamma_1,\dots,\gamma_p,0,-\gamma_p,\dots,-\gamma_1).
\end{equation}
The element-wise product between two bitstrings is denoted as $\bm{ab}$, and the inner product between two vectors as $\bm{\Gamma} \cdot \bm{a}$. 

We define the following function which isolates the dependency of the QAOA mixing angles
\begin{align}
f(\bm{a}) = \frac{1}{2} &\mel{a_1}{e^{i\beta_1 X}}{a_2}\mel{a_2}{e^{i\beta_2 X}}{a_3} \dots \mel{a_p}{e^{i\beta_p X}}{a_0} \\
&\mel{a_0}{e^{-i\beta_p X}}{a_{-p}} \dots \mel{a_{-3}}{e^{-i\beta_2 X}}{a_{-2}} \mel{a_{-2}}{e^{-i\beta_1 X}}{a_{-1}}. 
\end{align}
where we omit the dependency of $\bm{\beta}$ in the function argument to allow for compact notations. Factors of this form intuitively arise from the insertion of a resolution of the identity after the application of each QAOA layer which is indeed how these formulas are derived.

Finally, the recursive iteration that describes the contraction over the branches of the tree is given by
\begin{equation}  \label{eq:Hm_rec}
H_{d-1}^{(m)}(\bm{a}) = \left[ \sum_{\bm{b}} f(\bm{b}) H_{d-1}^{(m-1)}(\bm{b}) \exp(i \frac{1}{\sqrt{d}}  \bm{\Gamma} \cdot  \bm{a}\bm{b})
\exp(i \frac{h}{\sqrt{d}} \bm{\Gamma} \cdot  \bm{b}) \right]^{d-1},
\end{equation} 
with $m=1,\ldots, p$ and $H_{d-1}^{(0)}(\bm{a}) \equiv 1$. The level $m=1$ describes the contraction over the outermost leaves, while $m=p$ corresponds to the contraction level directly adjacent to the root(s).
Note however, that the subgraph of the local term has only a single root variable. Therefore, it has $d$ equivalent branches (and not just $d-1$), see also Fig.~\ref{fig:trees}. For this reason the final iteration, when $m=p$, needs to be modified in this case to
\begin{equation}   \label{eq:Hp_Z_rec}
H_{d}^{(p)}(\bm{a}) = \left[ \sum_{\bm{b}} f(\bm{b}) H_{d-1}^{(p-1)}(\bm{b}) \exp(i \frac{1}{\sqrt{d}}  \bm{\Gamma} \cdot  \bm{a}\bm{b})
\exp(i \frac{h}{\sqrt{d}} \bm{\Gamma} \cdot  \bm{b}) \right]^{d}.
\end{equation}

In Sec.~\ref{sec:results}, we will discuss our results that follow from minimizing Eq.~\eqref{eq:energy_density}. The time complexity of evaluating the energy density with the above described recursive procedure scales as $O((p+1) 2^{4p+2})$. In Appendix~\ref{sec:tree_calc} we derive (large) prefactor speed ups.

\section{MaxCut and Maximum independent set as Ising models} \label{sec:model}

In this section, we introduce two paradigmatic combinatorial optimization problems for which we will study the performance of QAOA in this work: Maximum Cut (MaxCut) and Maximum Independent Set (MIS). MaxCut refers to the task of finding a bipartition of the set of vertices of a given graph into two sets such that the number of edges of the graph connecting nodes between the two sets is maximal. The MIS problem asks for the largest set of vertices for a given graph such that no two vertices within this set are adjacent in the graph. The cost function for both MaxCut and MIS can be described as an Ising Hamiltonian of the form of Eq.~\eqref{eq:H_Ising} which enables us to treat them on a similar footing. We will consider both problems on $d$-regular graphs with $d\geq 3$. Note that for $d<3$, the MaxCut and MIS problems are equivalent and trivial.

\subsection{MaxCut}
When $h=0$ the model~\eqref{eq:H_Ising} corresponds to the standard Ising anti-ferromagnet, whose ground state encodes the solution of the MaxCut problem. Indeed, minimizing the energy when $h=0$, corresponds to maximizing the expected number of cut edges (or anti-ferromagnetic interactions)
\begin{equation}\label{eq:Maxcut}
C = \frac{1}{2} \sum_{ij \in E} (1 - Z_i Z_j).
\end{equation}

Suppose we have a QAOA ansatz state $\ket{\bm{\gamma},\bm{\beta}}$, obtained by minimizing the energy given by Eq.~\eqref{eq:H_Ising}, and we would like to assess the average quality of a candidate solution $\bm{z}$ sampled from this quantum state. For the MaxCut problem, finding a performance metric is straightforward, as every basis state $\bm{z}$ can be seen as a candidate solution, for which we can calculate the expected `cut fraction' $c\in \{ 0 , 1/|E|,\dots,1 \}$ with $|E|=Nd/2$,
\begin{equation}
c(\bm{z}) = \frac{1}{2|E|} \sum_{ij \in E} (1 - \ev{Z_i Z_j}{\bm{z}}).
\end{equation}
In the limit $N\rightarrow \infty$, the expected cut fraction for the depth-$p$ QAOA ansatz state minimizing the energy~\eqref{eq:H_Ising} is
\begin{equation}\label{eq:cp}
c_p = \lim_{N \rightarrow \infty} \frac{1}{2|E|} \sum_{ij \in E} (1 - \ev{Z_i Z_j}{\bm{\gamma},\bm{\beta}}) = \frac{1- \ev{Z_1 Z_2}{\bm{\gamma},\bm{\beta}}^{2\textrm{-tree}}}{2}.
\end{equation}
It was shown in Ref.~\cite{CojaOghlan2020} that there exist rigorous upper bounds $c_{ub}$ on the optimal cut fraction $c_{opt}$ ($c_{ub}>c_{opt}$) for $d$-regular graphs (and for sparse Erd\H{o}s-R\'{e}nyi graphs) in the limit $N\rightarrow \infty$. These bounds confirm the earlier conjecture presented in Ref.~\cite{Zdeborova2009} and can be obtained explicitly from techniques in statistical physics that make use of the interpolation method. In practice, they can be obtained by solving a small variational problem. We have calculated the upper bounds in this way and listed them in Table~\ref{tb:UBs} for different graph degrees.

Other upper bounds can be derived from complementary approaches, relating the improvement that is made on top of the average random sampling outcome to the Parisi constant~\cite{Dembo2015}. However, we have found that the first approach yields a tighter (i.e. lower) upper bound on the optimal fraction, at least for small graph degrees $d<10$. 

With the help of the tighter upper bound on the optimal cut fraction $c_{ub}$, we can compute a tighter (i.e. higher) lower bound $\alpha_{MC}$ on the approximation ratio $c_p/c_{opt}$, and thus on the performance, of QAOA for MaxCut 
\begin{equation}
c_p < \alpha_{MC} = \frac{c_p}{c_{ub}} <  \frac{c_p}{c_{opt}} \leq 1.
\end{equation}
In large parts of the QAOA literature, including the original QAOA paper, the expected cut fraction $c_p$ of the QAOA state is considered as a lower bound on the performance. While this is naturally correct (a weaker lower bound is still a lower bound), it may suggest the misleading insight that the performance of QAOA is worse for MaxCut on graphs with higher degree~\cite{Wurtz2020}. This is indeed wrong as in the extreme limit of (large) complete graphs, maximal cuts have size $|E|/2$, and such cuts are given on average by random sampling (which is equivalent to $p=0$ QAOA, or to QAOA with all angles 0). Hence, the MaxCut problem is easy in this limit, and the performance of QAOA, as measured by the above approximation ratio $\alpha_{MC}$, should therefore improve. 


\begin{table}[t]
\centering
\begin{tabular}{rcc}
\hline\hline
degree $d$ & $c_{ub}$ (cut fraction MaxCut) & $r_{ub}$ (independence ratio MIS) \\
\hline
3   & 0.92410 & 0.45400 \\
4   & 0.86824 & 0.41635 \\
5   & 0.83504 & 0.38443 \\
6   & 0.80500 & 0.35799 \\
7   & 0.78509 & 0.33567 \\
8   & 0.76585 & 0.31652 \\
9   & 0.75233 & 0.29987 \\
10  & 0.73877 & 0.28521 \\
20  & 0.67023 & 0.19732 \\
50  & 0.60820 & 0.11079 \\
100 & 0.57665 & 0.06787 \\
\hline\hline
\end{tabular}
\caption{The upper bounds on the optimal cut fraction $c_{opt}$ and independence ratio $r_{opt}$ for $d$-regular graphs in the asymptotic limit $N \rightarrow \infty$. For the MaxCut problem, these upper bounds have been obtained by solving the variational problem derived in Ref.~\cite{CojaOghlan2020}. For the MIS problem, the upper bounds have been taken directly from Ref.~\cite{Marino2020}, see also Refs.~\cite{McKay1987,Balogh2017}.}
\label{tb:UBs}
\end{table}

\subsection{Maximum independent set}
The MIS problem is about finding the largest subset of vertices of a given graph such that none of them is sharing an edge. The MIS problem is solved by finding the ground state of
\begin{equation}\label{eq:MIS}
H^{MIS}_{\lambda} = \lambda \sum_{ij \in E} N_i N_j - \sum_{i\in V} N_i 
\end{equation}
with $\lambda>1$, and the `number operator' $N_i = \frac{Z_i + 1}{2}$. Therefore $\ev{N_i}{z_i} = (z_i + 1)/2  \in \{ 0,1\}$. We assume that if $N_i \ket{z_i} = \ket{z_i}$, the $i$-th vertex is part of the chosen set. The second term in Eq.~\eqref{eq:MIS} is attempting to maximize the number of vertices in the set, i.e. the number of 1's in $\ket{\bm{z}}$, while the first term ensures their independence by adding an energy penalty for having two adjacent vertices in the set. For regular graphs, and in terms of Pauli-$Z$ operators, this Hamiltonian becomes 
\begin{equation} \label{eq:MIS_z}
H^{MIS}_{\lambda,d} = \frac{\lambda}{4} \sum_{ij \in E} Z_i Z_j  + \frac{\lambda d -2}{4} \sum_{i\in V} Z_i +  \frac{\lambda d N}{8} - \frac{N}{2}.
\end{equation}
After rescaling $H^{MIS}_{\lambda,d}\rightarrow 4/\lambda \, H^{MIS}_{\lambda,d}$, we see that this reduces to Eq.~\eqref{eq:H_Ising} with $h=d-2/\lambda$ up to an irrelevant constant. Hence, the ground state of Eq.~\eqref{eq:H_Ising} corresponds to the solution of the maximum independent set if $h \in \; ]d-2,d]$. However, the low-lying excited states might not correspond to independent sets. Note that for $h$'s outside this interval, even the ground state might not correspond to an independent set. For both reasons, a simple pruning protocol can be devised: if we have a candidate solution $\bm{z} \in \{-1,1\}^{\otimes N}$ with energy $H^{MIS}_{\lambda=1,d}(\bm{z}) = \ev{H^{MIS}_{\lambda=1,d}}{\bm{z}} < 0$, we can prune this bitstring to an independent set of at least size $-H^{MIS}_{\lambda=1,d}(\bm{z})$. Indeed, assume that $z_i=z_j=1$ and that $ij$ is an edge in $G$. If we remove this edge by flipping one of the variables randomly, the resulting state has at most the same energy $H^{MIS}_{\lambda=1,d}(\bm{z})$.
States with positive energy $H^{MIS}_{\lambda=1,d}(\bm{z}) > 0$ have less vertices chosen than edges existing that connect them. This implies that the above described pruning procedure might end up in a trivial independent set containing only one vertex. For states with positive energy, we can thus make the worst-case (trivial) assumption that they correspond to a single-node solution by keeping one of the 1's and flipping all others. Note, however, that the single node solution is `optimal' for complete graphs. Indeed, for complete graphs we have that the lower bound in the spectrum is $\min_{\bm{z}} H^{MIS}_{\lambda=1,d}(\bm{z})= -1$. 

Now, we would like to ask the same question as before for MaxCut: suppose we have a QAOA ansatz state $\ket{\bm{\gamma},\bm{\beta}}$, obtained by minimizing the energy given by Eq.~\eqref{eq:H_Ising}, what is the average quality of a candidate solution $\bm{z}$ sampled from this quantum state?
For the MIS problem, not every basis state $\bm{z}$ can be seen as a candidate solution, as not every basis state corresponds to an independent set. However, as discussed above, basis states with $H^{MIS}_{\lambda=1,d}(\bm{z}) < 0$ are easily prunable, to independent sets of at least size $-H^{MIS}_{\lambda=1,d}(\bm{z})$. The independence ratio is given by the number of variables in the independent set divided by the total number of variables. As performance metric for MIS, will thus consider
\begin{equation}
    r(\bm{z}) = 
- \frac{H^{MIS}_{\lambda=1,d}(\bm{z})}{N} =  - \frac{1}{4N} \sum_{ij \in E} \ev{Z_i Z_j}{\bm{z}}  + \frac{2-d}{4N} \sum_{i\in V} \ev{Z_i}{\bm{z}} + \frac{4-d}{8},  
\end{equation}
which is only meaningful when positive. In the asymptotic limit the expected independence ratio of bitstrings sampled from a QAOA state with negative energy $\ev{H^{MIS}_{\lambda=1,d}}$ is given by 
\begin{align}
r_p &= \lim_{N \rightarrow \infty} \left[ -\frac{1}{4N} \sum_{ij \in E} \ev{Z_i Z_j}{\bm{\gamma},\bm{\beta}} + \frac{2-d}{4N} \sum_{i\in V} \ev{Z_i}{\bm{\gamma},\bm{\beta}} +  \frac{4-d}{8} \right]\\
&= -\frac{d}{8} \ev{Z_i Z_j}{\bm{\gamma},\bm{\beta}}^{2\textrm{-tree}} + \frac{2-d}{4} \ev{Z_i}{\bm{\gamma},\bm{\beta}}^{1\textrm{-tree}} +  \frac{4-d}{8}.
\end{align}
Similarly, as for the cut fraction for the MaxCut problem, the optimal (largest) independence ratio for MIS on $d$-regular graphs $r_{opt}$ can be upper bounded for $N\rightarrow \infty$~\cite{McKay1987,Balogh2017}. So we can derive a $r_{ub} > r_{opt}$, which can then be used to form a lower bound $\alpha_{MIS}$ on the approximation ratio $r_p/r_{opt}$ achieved by QAOA on this problem in limit of large $N$
\begin{equation}
r_p < \alpha_{MIS} = \frac{r_p}{r_{ub}} <  \frac{r_p}{r_{opt}} \leq 1.
\end{equation}
The upper bounds $r_{ub}$ are also listed in Table~\ref{tb:UBs} for different degrees. These have been taken from Ref.~\cite{Marino2020}.

With the above introduced Ising model and performance metrics $\alpha_{MC}$ and $\alpha_{MIS}$, we will be able to investigate the performance of tree QAOA in solving the MaxCut and MIS problems as a function of the local field. From this section, it may seem intuitive that MIS problem is hard, as the magnitude of the local terms in Eq.~\eqref{eq:H_Ising} exactly balances the magnitude of the interactions. (Recall that the MIS regime is realized when $h\in]d-2,d]$.) Therefore, the MIS problem is realized exactly in the `critical' $h$ regime of Eq.~\eqref{eq:H_Ising}.

It has indeed been shown that the MIS problem exhibits a hardness property known as the `overlap gap property', and that local algorithms cannot find high quality solutions. We will review these topics in the next section.

\section{Local algorithms and overlap gap property} \label{sec:OGP}
Independent sets of more than half the size of the maximum independent set of random regular graphs above a certain degree are either very similar i.e. contain almost the same set of vertices, or very different i.e. contain almost disjoint sets of vertices~\cite{Gamarnik2013}. In other words, there is a gap in the overlap of large independent sets of graphs, which is why this phenomenon is called the Overlap Gap Property (OGP) of the MIS problem. Other NP-hard optimization problems like MaxCut are not believed to exhibit this phenomenon~\cite{Chen2017}. 
There is analytical evidence that this fundamental difference between the combinatorial optimization problems has a profound impact on the performance of local algorithms, classical and quantum alike~\cite{Chou2021}. 
Therefore, in this section we will briefly review $p$-local algorithms, and their performance limitations caused by OGP for the MIS problem.

An intuitive definition of generic $p$-local algorithms (be it quantum or classical) is given in Ref.~\cite{Chou2021}. We review this definition tailored to our purposes. 
We start by defining the $p$-local environment of a vertex $v_i \in V$ in the problem graph $G(V,E)$ as the subgraph $B^p_G(v_i) \subseteq G(V,E)$ induced by  vertex set $
 \{ v_j \in V \, | \, d_G(v_i,v_j) \leq p \}.
$
Here the graph distance between two vertices $d_G(v_i,v_j)$ is the number of edges in the shortest path between $v_i$ and $v_j$.

We consider a probabilistic algorithm $\mathcal{A}$ that generates samples from a distribution $\mathcal{A}_G(z_1,..,z_N)$ of binary variables $z_1, \dots, z_N$ that are associated to each vertex of a specific graph $G$. We say that $\mathcal{A}$ is a $p$-local algorithm when it satisfies two criteria: 
\begin{enumerate}[label=(\roman*)]
\item When $d(v_i,v_j) > 2p$, $\mathcal{A}_G(z_i)=\sum_{z_1, \dots, z_N \setminus z_i}\mathcal{A}_G(z_1,..,z_N)$ is statistically independent from $\mathcal{A}_G(z_j)=\sum_{z_1, \dots, z_N \setminus z_j}\mathcal{A}_G(z_1,..,z_N)$, meaning that the marginals are statistically independent if the vertices associated to the variables do not have overlapping environments.
\item The marginals are equal $\mathcal{A}_G(z_i)$ = $\mathcal{A}_G(z_j)$ if their $p$-local environments are isomorphic, $B_G^p(v_i) \simeq  B_G^p(v_j)$. 
\end{enumerate}

In this sense QAOA is a local algorithm because (i) Variables with non-overlapping RCCs are not mutually entangled and also not classically correlated. Hence, they do not share any mutual information and are therefore statistically independent. (ii) Observables depend solely on the topology of their RCCs for a given set of angles.

We now review the definition of the OGP tailored to the MIS problem on regular graphs. Further details can be found in Refs.~\cite{Gamarnik2013, Farhi2020a}. We start by considering a multiplicative factor $\mu\in [0,1]$, that quantifies the distance between a candidate solution and the optimum. For a given graph $G$, we can thus define the set of $\mu$-good solutions (where we will assume that the bitstring $\bm{z}$ corresponds to an independent set of $G$)
\begin{equation} 
    \mathcal{S}(\mu,G) = \{ \bm{z} : r(\bm{z}) \geq \mu r_{opt} \}.
\end{equation}
Now let us consider two random regular graph instances $G_0$ and $G_1$, defined on $N$ vertices, each with their corresponding sets of $\mu$-good solutions. Let us define for every $0< \theta \leq \mu $, two sets of pairs $(\bm{z}_0, \bm{z}_1)$ with $\bm{z}_i \in \mathcal{S}(\mu,G_i) $, $i=0,1$, such that
\begin{equation}
    \mathcal{\mathcal{S}}^{\textrm{similar}}(\mu,\theta,G_0,G_1) = \{(\bm{z}_0, \bm{z}_1)  : r(\bm{z}_0 \land \bm{z}_1) \geq \theta r_{opt} \},
\end{equation}
and
\begin{equation}
    \mathcal{\mathcal{S}}^{\textrm{different}}(\mu,\theta,G_0,G_1) = \{(\bm{z}_0, \bm{z}_1)  : r(\bm{z}_0 \land \bm{z}_1) \leq \theta r_{opt} \}.
\end{equation}
These sets thus contain pairs of good solutions whose independence ratio of the intersection normalized to the optimal independence ratio, is either equal, or smaller or bigger than $\theta$. Here the intersection of two independent sets means taking the logical \texttt{AND} product between their bitstrings, i.e. element-wise $z_i \land z_j = 1$ if and only if $z_i=1$ and $z_j=1$.

We now define interpolation graphs $G_t$ between the two random regular graphs $G_0$ and $G_1$. Here $t\in \{0,1/|E|,\dots,1\}$ determines the fraction of edges that are selected from $G_1$ and added to $G_t$. At the same time a fraction of $(1-t)$ edges are selected from $G_0$ and added to $G_t$. To have similarity between $G_t$ and $G_{t+\Delta}$ with $\Delta=|E|^{-1}$, the previously added (removed) edges from $G_1$ ($G_0$) should remain the same, such that $G_t$ and $G_{t+\Delta}$ only differ by two edges in the typical case. For the sake of stating the OGP property of MIS, it does not matter that $G_t$ is not regular at every step in the interpolation. The average degree of $G_t$ is still $d$, which is the important fact. 

\begin{thm}\label{thm:OGP}
    See Ref.~\cite{Farhi2020}. The MIS problem on $d$-regular graphs has OGP when $d$ is large enough. This means that there exist a $\mu^{\star}$ such that for every $\mu > \mu^{\star} $ there exists $0<\theta_1<\theta_2<\mu$, such that for $N$ large enough
\begin{equation}
    \mathcal{\mathcal{S}}^{\textrm{similar}}(\mu,\theta_1,G_{t_1},G_{t_2}) \cap \mathcal{\mathcal{S}}^{\textrm{different}}(\mu,\theta_2,G_{t_1},G_{t_2}) = \emptyset, \quad \forall \, 0 \leq t_1,t_2 \leq 1   
\end{equation}
with high probability (i.e. a probability converging to 1 exponentially fast in $N$), and
\begin{equation}
    \mathcal{\mathcal{S}}^{\textrm{similar}}(\mu,\theta_1,G_{0},G_{1}) = \emptyset  
\end{equation}
with high probability. 
\end{thm}

Hence, the OGP means that for bounded degree random graph instances $G_{t_1}$ and $G_{t_2}$, pairs of $\mu$-good solutions are either `similar' (they have at least normalized intersection ratios of $\theta_2$), or `different' (they have at most normalized intersection ratios of $\theta_1$) with high probability. Pairs of $\mu$-good solutions that do not satisfy this property are exponentially rare. Additionally, when two graph instances do not share an a priori similarity, like $G_0$ and $G_1$, it is already exponentially rare to have pairs of similar $\mu$-good solutions with normalized intersection ratios larger than $\theta_1$. This means that somewhere in the interpolation between $G_0$ and $G_1$, there must be a `jump' allowing for pairs of similar solutions.

It is shown in Ref.~\cite{Gamarnik2013} that the presence of OGP for MIS on regular graphs with large $d$ obstructs local algorithms, like $p$-local QAOA~\cite{Farhi2020a,Farhi2020}, from finding $(\mu > \frac{1}{2} + \frac{1}{2\sqrt{2}})$-good solutions with non-vanishing probability. We briefly sketch why. Let us construct a sequence of coupled independent sets for the interpolation graphs $G_t$, which we call $\bm{z}_t$. This sequence starts from an independent set $\bm{z}_0$ obtained by taking a single sample from running the depth-$p$ QAOA algorithm for $G_0$, potentially followed by pruning. QAOA exhibits strong Hamming weight concentration~\cite{Farhi2020a}. Therefore, $r(\bm{z}_0)$ will not differ much from its average $r_p^0$ with high probability. Because we know that the next graph $G_{\Delta}$ only differs by at most two edges from $G_0$, we can construct an independent set $\bm{z}_{\Delta}$ that is similar to $\bm{z}_0$. Indeed, as $G_{\Delta}$ only differs by at most two edges from $G_0$, there exist QAOA samples for $G_{\Delta}$ that will only be different in at most $4\max_i |B_G^p(v_i)|$ bits from $\bm{z}_0$. We take  $\bm{z}_{\Delta}$ to be one of those. Notice that for fixed $p$ and large $N$, the subgraph sizes are vanishing compared to $N$, so that is why $\bm{z}_0$ and $\bm{z}_{\Delta}$ are considered similar. Like this we can construct the sequence of similar solutions $\bm{z}_t$. 

We now assume that depth-$p$ QAOA can create large independent sets of $G_t$
\begin{equation} \label{eq:assumption_QAOA_good}
    r_p^t > \mu r_{opt}.
\end{equation}
Notice that concentration over graph instances~\cite{Makover2006} implies that assumption~\eqref{eq:assumption_QAOA_good} must be valid $\forall \,t$. However this assumption is in direct violation with the OGP of MIS, see Theorem~\ref{thm:OGP}, because it would then be possible with high probability to have similar large independent sets, for which $\bm{z}_0 \land \bm{z}_t$ is large, without having a sudden jump to different large independent sets, for which $\bm{z}_0 \land \bm{z}_{t+\Delta}$ is small, for a certain $t$. Therefore, we conclude that the expected independence ratio obtained from QAOA is suboptimal for large enough $N$ and $d$
\begin{equation} 
    r_p \leq \mu r_{opt}.
\end{equation}

 It is shown in Ref.~\cite{Rahman2014} that asymptotically $\mu^{\star} \xrightarrow[]{>} 1/2$, for $d \rightarrow \infty$. Hence, there is a performance gap that can be as large as $1/2$ between the output of local algorithms and optimality. (Famously, this is also true for Erd\H{o}s-R\'{e}nyi graphs that are both sparse and dense~\cite{CojaOghlan2010,Wein2020}.) Indeed, it is shown in Ref.~\cite{Frieze1992} that the optimal (i.e. maximum) independence ratio asymptotically converges to $2\log(d)/d$, for $d \rightarrow \infty$. However, it is shown in Ref.~\cite{Rahman2014} that the largest independence ratios that can be obtained by local algorithms are asymptotically $\log(d)/d$, when $d \rightarrow \infty$.


\begin{table}[t]
\centering
\begin{tabular}{rc}
\hline\hline
\textbf{degree $d$} & \textbf{Lower bound on $\mu^{\star}$ for MIS} \\
\hline
3   & 0.9809 \\
4   & 0.9705 \\
5   & 0.9346 \\
6   & 0.9300 \\
7   & 0.9255 \\
8   & 0.9098 \\
9   & 0.9057 \\
10  & 0.9021 \\
20  & 0.8808 \\
50  & 0.8583 \\
100 & 0.8427 \\
\hline\hline
\end{tabular}
\caption{The lower bounds on the possible onset of OGP for the MIS problem for $d$-regular graphs in the asymptotic limit $N \rightarrow \infty$. These bounds are the approximation ratios achieved by the linear-time classical algorithm presented in Ref.~\cite{Marino2020}, which is the state-of-the-art classical local polynomial-time algorithm solving this problem.}
\label{tb:ogp_mis_bounds}
\end{table}

 These findings refuted the hope~\cite{Hatami2012} that local algorithms may be able to find maximum independent sets in random regular graphs. However, the statements are only proven for `sufficiently large' $d$, which leaves open the question if OGP holds for the smallest $d$, and if it holds, what would be the value of $\mu^{\star}$. From Ref.~\cite{Marino2020}, some lower bounds for the possible onset of OGP can however be deduced, see Table~\ref{tb:ogp_mis_bounds}. Hence, the degrees considered in this paper seem still rather far from the asymptotic $d$ regime. This also implies that even if there is OGP existing for the smallest $d$, the regime in which it exists is small. Indeed, for instance for $d=3$, Table~\ref{tb:ogp_mis_bounds} means that there exists a local algorithm that can create independent sets with ratios of at least $0.9809 r_{opt}$. Hence if the MIS problem for $d=3$ would have an OGP, $\mu^{\star}$ must be bigger than this value. In other words, local algorithms fail the earliest in creating independent sets with ratios larger $0.9809 r_{opt}$ if the MIS for $d=3$ problem would have an OGP.

In general, a proven presence of OGP for a problem class, yields a limitation on the performance of \textit{all} local algorithms~\cite{Chou2021,Gamarnik2021}. On the other hand, it is a standing conjecture that the MaxCut (and Sherrington-Kirkpatrick) problems do not have OGP~\cite{Chen2017,Auffinger2017, Chou2021}, and that there is thus no such performance gap set by a certain $\mu^{\star}$ for local algorithms. Indeed, under the assumption that these problems have no OGP, approximate message-passing algorithms have been devised that with high probability do not see such a gap~\cite{Montanari2018,Alaoui2021}. In Ref.~\cite{Alaoui2021} in particular, such an algorithm for the MaxCut problem is constructed that finds solutions that are $(1-\epsilon)$ close to optimality with high probability. It relies however on the condition that $d>O(1/\epsilon)$. Hence $\epsilon$ can not be made arbitrarily small yet. 

The conjecture of `no OGP' for MaxCut suggests that there are no bounds on the expected performance of tree QAOA in the asymptotic limit when $p$ is increased. On the other hand, for MIS the regime with OGP can certainly not be reached with $p$-local QAOA when the RCCs do not span the whole system~\cite{Farhi2020a,Farhi2020}. The tree QAOA, which is the subject of this paper, however considers the asymptotic limit $N\rightarrow \infty$ and fixed $p$. In that case, it is also expected that even when increasing $p$ indefinitely, the performance of tree QAOA is bounded away from optimality~\cite{Goh2024}. 

In the next section, we will evaluate the performance of tree QAOA explicitly for both problems.

\section{Tree QAOA results} \label{sec:results}
In this section, we first present the results from numerically evaluating the recursive tree QAOA formulas [see Eqs.~\eqref{eq:Z_rec},\eqref{eq:ZZ_rec},\eqref{eq:Hm_rec} and~\eqref{eq:Hp_Z_rec}] in order to minimize the Ising energy density Eq.~\eqref{eq:energy_density}. These formulas were summarized in their simplest form in Sec.~\ref{sec:QAOA}, and are derived (and made faster, by prefactors) in Appendix~\ref{sec:tree_calc}. Second, we recycle the angles obtained from the tree QAOA to solve finite-size instances. In this way, we see that for degree $d=3,4$, we outperform the GW algorithm on average for MaxCut with those fixed angles~\cite{Wurtz2020,Wurtz2021}. For MIS, we outperform a minimal greedy algorithm for $d=3$.

\subsection{Performance of tree QAOA in the asymptotic limit}

\begin{figure}
    \centering
     \includegraphics[width=0.6\textwidth]{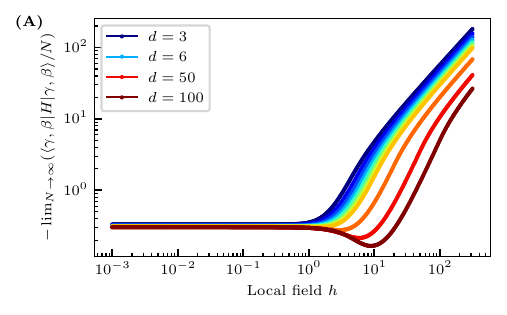}
    \includegraphics[width=0.6\textwidth]{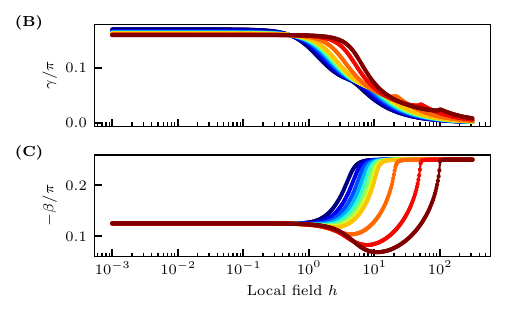}
    \caption{\textbf{(A) The asymptotic energy densities} achieved by $p=1$ QAOA as a function of the local field [see Eq.~\eqref{eq:energy_density}]. The colors indicate the different graph degrees listed in Table~\ref{tb:UBs}, only a selection is shown in the legend. \textbf{(B,C) The respective QAOA angles to obtain these energy densities.}}
    \label{fig:p1_energy}
\end{figure}

As a first step, we will optimize the QAOA angles such that the energy density Eq.~\eqref{eq:energy_density} is minimized as a function of the local field. Secondly, we will evaluate the performance metrics presented in Sec.~\ref{sec:model} on the resulting ansatz state for both problems. Here, the operator in the phase separator of the QAOA ansatz is the same as the objective function that is minimized (i.e. Eq.~\eqref{eq:H_Ising}), but is not necessarily equivalent to the performance metric. Indeed for MaxCut they are only directly related when $h=0$, and for MIS when $h=d-2$. Later we will however restrict to these cases. We first investigate the (fixed) performance metrics for both problems as a function of the local field because in that way we can view the performance for different degrees and fields in a unified way. Like that, we can also investigate the QAOA performance of MaxCut in the $h$-regime corresponding to MIS, and conversely. Such analysis leads to interesting observations like the fact that the performance of QAOA in solving the MIS problem does not vary much with $h$ when $h<d$ and both $d$ and $p$ are small.  

\subsubsection{Changing the local field}

In Fig.~\ref{fig:p1_energy}{(A)}, we show the resulting energy density Eq.~\eqref{eq:energy_density} as a function of the local field for $p=1$. The corresponding QAOA angles to obtain these energy densities are shown in Fig.~\ref{fig:p1_energy}{(B,C)}. The behavior of these is quite intuitive: for small $h<1$ there is an anti-ferromagnetic regime (constant non-trivial $\gamma$), for $1<h<d$ a transitional `critical' regime (changing non-trivial angles), and for $d<h$ a trivial regime where all spins simply anti-align with the local field (realized exactly by the $\exp(i\pi/4 \sum_i X_i)$ rotation, and vanishing $\gamma$).  

Next, we evaluate the performance of these optimized QAOA ans\"{a}tze in solving the MaxCut and MIS problems. The approximation ratio we evaluate as performance metric for MIS (see Sec.~\ref{sec:model}) can be seen as an order parameter that detects `critical order' in the QAOA state that is optimized for $H$ as a function of $h$. Similarly, the MaxCut approximation ratio simply detects anti-ferromagnetic order.

\begin{figure}
    \centering
     \includegraphics[width=0.6\textwidth]{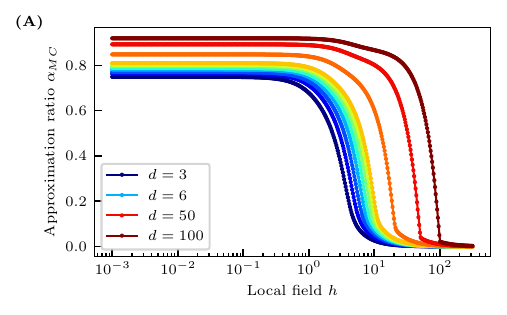}
    \includegraphics[width=0.6\textwidth]{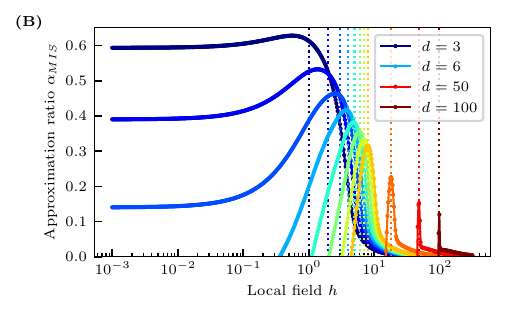}
    \caption{\textbf{The approximation ratios for (A) MaxCut and (B) MIS} obtained by $p=1$ QAOA optimized as a function of the local field for $N\rightarrow \infty$. The colors indicate the different graph degrees listed in Table~\ref{tb:UBs}, only a selection is shown in the legend. The dotted vertical lines indicate local fields $h=d-2$. The QAOA angles are chosen such that the energy density given by Eq.~\eqref{eq:energy_density} is minimized [see Fig.~\ref{fig:p1_energy}]. The approximation ratios $\alpha_{MC}$ and $\alpha_{MIS}$ as discussed in Sec.~\ref{sec:model}, are evaluated on the resulting QAOA state.  }
    \label{fig:p1_performance}
\end{figure}

 In Fig.~\ref{fig:p1_performance}, we show the corresponding lower bounds on the QAOA approximation ratios for the two problems as a function of the local field in the limit $N\rightarrow \infty$. We recall that these approximation ratios take into account the state-of-the-art asymptotic upper bounds listed in Table~\ref{tb:UBs}. We also recall that a zero approximation ratio would mean cutting no edges for MaxCut, or finding a vanishing independence ratio with $N$ for MIS. The behaviour of these approximation ratios is interesting. First, we can observe the manifestly different problem hardness when $d$ grows larger in the performance of QAOA for both problems. On the one hand, MaxCut in the large $d$ limit becomes `easier'. Indeed, when approaching complete graphs, cutting half of the edges (or random sampling), becomes optimal. However, on the other hand MIS is highly non trivial as it has a proven OGP when $d$ grows larger~\cite{Gamarnik2013}. This is reflected in a vanishing performance of QAOA when $d$ is increased. Interestingly, for MIS on 3-regular graphs, the approximation ratio achieved by $p=1$ QAOA is surprisingly constant when $h<3$ [see Fig.~\ref{fig:p1_performance}{(B)}]. This implies that solutions for MIS on 3-regular graphs with independence ratios that are $\sim60\%$ of the optimum can be equally well sampled from a QAOA circuit with angles optimized for the MaxCut problem. This constant behavior is specific to low degrees and relatively low approximation ratios, and is expected to vanish with increasing $d$ (or $p$).

\begin{figure}
    \centering
     \includegraphics[width=0.6\textwidth]{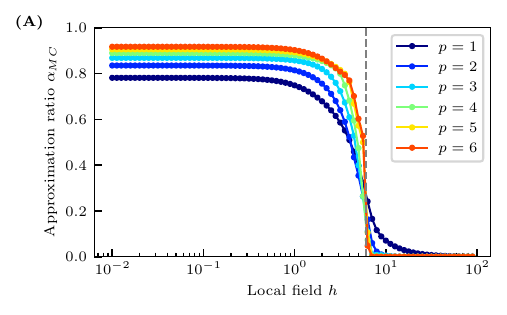}
    \includegraphics[width=0.6\textwidth]{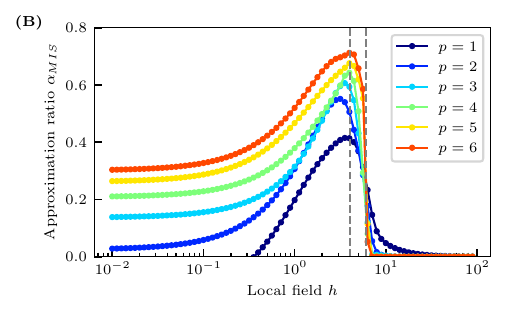}
    \caption{\textbf{The approximation ratios obtained by QAOA for (A) MaxCut, and (B) MIS on 6-regular graphs} for various $p$ in the limit $N \rightarrow \infty$. The QAOA ansatz is optimized to minimize the energy given by Eq.~\eqref{eq:H_Ising}. Therefore, there is a sharp transition in solution quality at $h=d$ for both problems. The ground state of Eq.~\eqref{eq:H_Ising} corresponds to the solution of MaxCut when $h\ll d$, and to the solution of MIS when $h\in]d-2,d]$. In these indicated regions the approximation ratio must therefore increase with $p$.}
    \label{fig:fixed_d_performance}
\end{figure}

There is a sharp transition in the achieved approximation ratios for both problems when $h/d>1$. This transition is due to the fact that the trivial regime is entered where all spins simply anti-align with $h$, leading to a diamagnetic state that represents an empty set and that cuts no edges. As QAOA capures this behavior more accurately with increasing $p$, this transition gets also sharper with increasing $p$. This can be seen from Fig.~\ref{fig:fixed_d_performance}, where $d=6$ is fixed, and $p$ increased. 

 For the MIS case, the performance vanishes when $h$ falls below some threshold value if $d$ is large enough. From Figs.~\ref{fig:p1_performance}{(B)} and \ref{fig:fixed_d_performance}{(B)}, it can be seen that this happens when $d=6$ for $p=1$. Increasing $h$ towards $d-2$ resolves this behavior, but notice that also increasing $p$ helps to resolve this behavior. For $d=6$, taking $p=2$ seems already sufficient [see Fig.~\ref{fig:fixed_d_performance}{(B)}]. 

As expected from the problem definition in Sec.~\ref{sec:model}, the MIS approximation ratio reaches a peak when $h$ is around $d$, marking the critical regime of the Ising model Eq.~\eqref{eq:H_Ising} and corresponding to a positive independence constraint. Interestingly, however, for small $p$ the maximal performance seems to be reached when $h\lesssim d-2$ corresponding to an independence constraint $\lambda \lesssim 1$. This indicates that for QAOA with small $p$, it is not a requirement that $h$ is chosen in such a way that the ground state of Eq.~\eqref{eq:H_Ising} corresponds to the maximum independent set. However, when $p$ is increased and when therefore the energy distribution of QAOA shifts towards lower energies, the peak in optimal performance shifts towards a field that corresponds to an independence constraint $\lambda > 1$. This behavior can be seen in Fig.~\ref{fig:fixed_d_performance}{(B)}, where the boundaries of the region with independence constraint $\lambda > 1$ are indicated by the two dashed lines.

\subsubsection{Fixing the local field}

\begin{figure}
    \centering
     \includegraphics[width=0.6\textwidth]{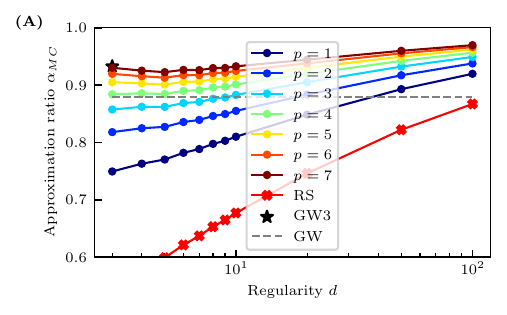}
    \includegraphics[width=0.6\textwidth]{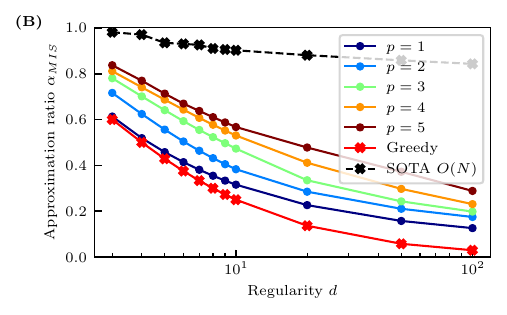}
    \caption{\textbf{The approximation ratios for (A) MaxCut and (B) MIS} obtained by depth-$p$ QAOA for $N \rightarrow \infty$. For MaxCut, we compare with the GW guarantee and Random Sampling (RS). For 3-regular graphs there is a specialized version of the GW algorithm, discussed in Ref.~\cite{Halperin2004}, achieving a worst-case approximation ratio of $0.9326$ which is indicated by the star. For MIS, we compare to the performance guarantee of a minimal greedy search~\cite{Halldrsson1997}, and the state-of-the-art linear-time prioritized search algorithm of Ref.~\cite{Marino2020}. }
    \label{fig:fixed_h_performance}
\end{figure}

Now, we fix the local field to $h=0$ for MaxCut and to $h=d-2$ for MIS (corresponding to $\lambda=1$), and investigate the performance of tree QAOA with increasing $p$ for different graph degrees. For MaxCut the performance of tree QAOA has been investigated in the literature before~\cite{Streif2020,Wurtz2020,Wurtz2021,Basso2021}, it even has been considered for $p=1$ in the original QAOA paper~\cite{Farhi2014}. However, our current work extends these analyses by: (i) further improving the achieved approximation ratios by employing the upper bounds as discussed in Sec.~\ref{sec:model}, and (ii) extending the analysis to degrees larger than $d=3$. To make the first point explicit, it is stated in the original QAOA paper that on 3 regular graphs, QAOA is guaranteed to find approximation ratios of at least $\sim 0.69$ (also for $N\rightarrow \infty$). By incorporating the upper bound on the cut fraction of 3-regular graphs, we realize the approximation ratio is actually at least $\sim 0.75$ for $N\rightarrow \infty$, see Fig.~\ref{fig:fixed_h_performance}{(A)}. Here, we compare the improved approximation ratios obtained by QAOA, with the guaranteed approximation ratio of the Goemans-Williamson (GW) algorithm~\cite{Goemans1995} and random sampling. The GW algorithm is guaranteed to find a solution to the MaxCut problem with an approximation ratio of at least $\alpha_{GW} = \min_{x\in [0,1]} 2\arccos(x)/(\pi (1-x)) = 0.87\dots$ for any instance. Indeed, denoting the cut fraction of an average GW solution as $c_{GW}$, we have that
\begin{equation}
\alpha_{GW} \, c_{opt}  \leq \alpha_{GW} \, c_{sdp} \leq  c_{GW},  
\end{equation}
where $c_{opt}$ is the cut fraction corresponding to the maximum cut, and $c_{sdp}$ the (optimal) solution of the semi-definite program following the relaxation of the binary variables to unit vectors on the $N$-dimensional sphere. The first inequality directly follows from the fact this solution provides an upper bound on $c_{opt}$, i.e. $c_{opt} \leq c_{sdp}$, because it is a relaxation. The second inequality follows from the fact that the GW rounding produces a cut whose size is on average at least a fraction $\alpha_{GW}$ of the semi-definite relaxation optimum.

Thus, the knowledge of a tighter upper bound on $c_{opt}$ does not alter the performance guarantee $\alpha_{GW}$ of the GW algorithm. So the approximation ratios introduced in Sec.~\ref{sec:model} compare directly to $\alpha_{GW}$, and already pass $\alpha_{GW}$ at $p=4$ for the lowest degrees. However, we expect that the practically achieved GW approximation ratios will improve with increasing $d$. In particular, the GW results will never fall below random sampling $(2c_{ub})^{-1}$ shown by the red line in Fig.~\ref{fig:fixed_h_performance}{(A)}. On the other hand, for small $d$, we already pass $\alpha_{GW}$ at $p=4$ which leaves open the possibility that  shallower circuits than previously assumed could already provide a performance and runtime advantage over the GW algorithm. 

We also note that in Fig.~\ref{fig:fixed_h_performance}{(A)}, the gap in approximation ratio between random sampling and QAOA shrinks upon increasing $d$. Based on this, it is also an interesting question of how QAOA performs relative to random sampling. For this, we can easily define a modified approximation ratio resembling the Q-score ratio defined in Ref.~\cite{Martiel2021} (see Appendix~\ref{ssec:relative_maxcut}). Then, we can conclude that similarly to the local classical algorithm of Ref.~\cite{Hirvonen2014}, QAOA achieves constant approximation ratios relative to random sampling in the large $d$ limit. This finding is consistent with the observation of a vanishing gap between random sampling and QAOA in Fig.~\ref{fig:fixed_h_performance}{(A)}. We also determine the constant explicitly in Appendix~\ref{ssec:large_d} for $p=1$ and conclude that 
\begin{equation}
    \alpha^{\mathrm{relative}}_{MC}(d\rightarrow \infty, p=1) = \frac{1}{2\sqrt{e} P^{\star}} \approx 0.397\dots,
\end{equation}
meaning that QAOA with $p=1$ is around $40\%$ better than random sampling in the large $d$ limit.

Now, we analyze the performance of tree QAOA for solving MIS problem. We show the obtained approximation ratios in Fig.~\ref{fig:fixed_h_performance}{(B)}. We compare to the performance guarantee of the minimal greedy algorithm. The minimal greedy algorithm makes locally minimal random choices. It consists out of the following steps: (i) Randomly select a vertex from $G$ that has the lowest degree. (ii) Add this vertex to the independent set, and delete all its neighbors from $G$. (iii) Repeat (i) and (ii) until the the graph $G$ is empty (i.e. has no edges left). (iv) Add possible remaining vertices to the independent set. This algorithm is guaranteed to find independent sets with approximation ratios $3/(d+2)$~\cite{Halldrsson1997}. We also compare to the best known classical linear-time algorithm of Ref.~\cite{Marino2020}, see dashed grey line. (These approximation ratios are the values that we considered as a lower bound for $\mu^{\star}$ in Table~\ref{tb:ogp_mis_bounds}.) As can be seen, there is an increasing gap with $d$ between the approximation reached by QAOA and this algorithm. Therefore, it seems that the performance of tree QAOA for MIS is vanishing upon increasing $d$. This observation can be made explicit for $p=1$ [see Appendix ~\ref{ssec:large_d}]. We would expect this to be true as well when $p>1$ and finite, because in the limit $N\rightarrow \infty$ we would expect that the locality restrictions apply in the same way for $p>1$ as for $p=1$.  
In practice, we are however mostly interested in solving the problems for small and fixed $d$. Then, the optimal parameters obtained from the tree QAOA could also be used for small problems, of problem size corresponding to the currently available QPU's. Illustrating that this approach achieves good performances is the topic of the next section. 

\subsection{Tree QAOA angles applied to finite-size problems}

\begin{figure}
    \centering
     \includegraphics[width=0.6\textwidth]{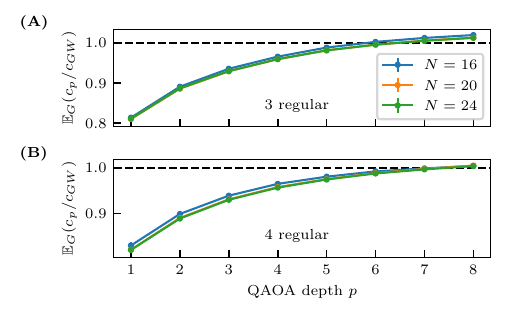}
    \includegraphics[width=0.6\textwidth]{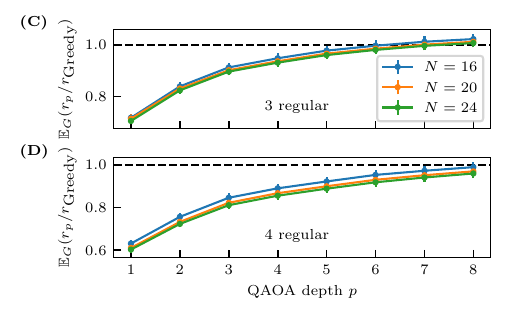}
    \caption{\textbf{Performance comparison between QAOA with predetermined tree angles and classical algorithms} on small instances of 3 and 4 regular graphs. The tree angles have been determined by the techniques outlined in Appendix~\ref{sec:tree_calc}. \textbf{(A,B) MaxCut QAOA compared to GW}. The comparison is made on an instance-by-instance basis: we compare the expectation value of the number of cut edges measured on the QAOA state with the average number of cut edges based on 100 GW runs. \textbf{(C,D) MIS QAOA compared to minimal greedy}. We sampled 200 random graph instances, and the error bars show $3\bar{\sigma}$ with $\bar{\sigma}$ the standard error of the mean. The QAOA angles used for these simulations are included in Appendix~\ref{sec:angles_used}.}
    \label{fig:finite_size}
\end{figure}

The goal of this section is to compare how well QAOA with the tree angles obtained in the previous section performs for finite-size instances. In the previous section [see Fig.~\ref{fig:fixed_h_performance}], we compared the expected QAOA outcomes with the performance guarantees of classical algorithms: GW for MaxCut and minimal greedy for MIS. However, it may be that in practice these classical algorithms perform significantly better than their lower performance bound. At the same time, the tree angles are only guaranteed to be optimal in the asymptotic limit. However, for MaxCut, it has been observed before that these angles also work well away from the asymptotic limit~\cite{Streif2020,Wurtz2021}.

To shine a clearer light on these issues, our goal is to make an explicit comparison for both MaxCut and MIS to the classical algorithms. The protocol we use is the following: (i) Use a set of tree angles determined as before (by making use of the formulas in Appendix~\ref{sec:tree_calc}) to create a QAOA ansatz state for the sampled finite-size problem instances $G(N,d)$. Hence, we do not apply a procedure to fine tune the QAOA angles, but use the same set of fixed angles for all instances. (ii) Measure the performance metrics, i.e. the expectation values of Sec.~\ref{sec:model}, on this QAOA ansatz state. (iii) Run a classical probabilistic algorithm on the same instance a certain number of times, and take the average. (iv) Compare the outcomes of both algorithms. 

For MaxCut, we set $h=0$ in Eq.~\eqref{eq:H_Ising} and compare the fixed angle QAOA to the GW algorithm~\cite{Goemans1995}. We measure the expected number of cut edges in the prepared QAOA state for every instance. For each instance, we then run the GW algorithm 100 times, and take the average. We observed that in practice, in our setup, the GW performs on average significantly better than its worst-case guarantee $\alpha_{GW}$. This also explains why we only observe an improvement for $p=8$ for 3 and 4 regular graphs, see Fig.~\ref{fig:finite_size}{(A,B)}. This is in contrast to the asymptotic limit, where tree QAOA already outperforms the \textit{lower bound} at $p=4$ [see Fig.~\ref{fig:fixed_h_performance}]. Although the approximation ratios shown in Fig.~\ref{fig:fixed_h_performance} clearly increase when growing $d$, we observe for higher $p$ a non-monotonic regime when $d$ is small. This possibly explains the slightly worse performance for $d=4$ compared to $d=3$ in Fig.~\ref{fig:finite_size}{(A,B).

For MIS, we set $h=d-2$ in Eq.~\eqref{eq:H_Ising} and compare to the minimal greedy algorithm. Such greedy algorithms are widely used algorithms, but their performance is far from optimal for\textit{ large graphs}, even among local algorithms. Indeed, from Fig.~\ref{fig:fixed_h_performance}{(B)}, it is clear that we are far from outperforming the state-of-the-art (SOTA) classical local linear-time algorithm that solves the MIS problem on large regular graphs~\cite{Marino2020}. However, in the extreme \textit{finite-size scenario}, we actually found that the minimal greedy algorithm outperforms the SOTA algorithm which justifies the choice of using it for this comparison here.  In addition, it has a straightforward implementation and a performance guarantee. This comparison is shown in Fig.~\ref{fig:finite_size}{(C,D)} for 3 and 4 regular graphs. We see that tree QAOA is outperforming the minimal greedy algorithm for $p=8$ in the case $d=3$, however this is not anymore the case for $d=4$. First, these results confirm that, similarly to the GW algorithm, also the greedy algorithm performs significantly better on average than its worst-case guarantee. Second, these results for QAOA on small system sizes seem to also reflect the $N\rightarrow\infty$ finding that the MIS problem becomes more challenging for QAOA as $d$ increases. 

The tree angles we have used for the simulations in this section are included in Appendix~\ref{sec:angles_used}.

\section{Conclusion} \label{sec:concl}
In this work, we have provided a comprehensive analysis of the performance landscape of tree QAOA for the MaxCut and MIS problems on
random $d$-regular graph by unifying them as Ising models. Our approach extends the recursive formula technique introduced in Ref.~\cite{Basso2021} to handle Ising models with a local field, allowing us to evaluate the QAOA performance for both problems for large but finite graph degrees up to $d=100$. We combined these results with state-of-the-art upper bounds on optimality which led to our main result: the MaxCut and MIS approximation ratios achieved by QAOA in the limit $N\rightarrow \infty$ shown in Fig.~\ref{fig:fixed_h_performance} for many different graph degrees $d$. 

Our analysis demonstrates that QAOA outperforms the Goemans-Williamson (GW) guarantee at much shallower depths than previously assumed. Specifically, we show that asymptotic tree QAOA can surpass GW for MaxCut when $p \geq 4$ in the large graph limit ($N\rightarrow \infty$), rather than requiring a depth of $p \geq 11$~\cite{Wurtz2021}. 

For the MIS problem, QAOA at any depth $p\geq 1$ outperforms the lower bound of the greedy algorithm. However, QAOA is far from outperforming the best classical local polynomial-time algorithm for this problem~\cite{Marino2020}. For higher degree, the performance of QAOA with fixed $p$ is bounded away from optimality for the MIS problem due to the presence of OGP~\cite{Farhi2020a,Farhi2020}. The increased hardness of the MIS problem with increasing $d$ is clearly manifested by the decrease of approximation ratios achieved by tree QAOA [see Fig.~\ref{fig:fixed_h_performance}]. This behavior is opposite for MaxCut which is believed not to exhibit an OGP. In this case the QAOA approximation ratios are indeed not vanishing with increasing $d$.

In addition, we also showed that QAOA has a promising performance when the tree angles are taken as `fixed angles' for small problems. For MaxCut, this setting is similar as in Ref.~\cite{Wurtz2020} and we showed that QAOA with fixed angles outperforms the GW algorithm at $p=8$ for $d=3,4$ when comparing on average to explicit GW runs. Such an analysis seems to also partly hold for the MIS problem when comparing to minimal greedy at low degree. Indeed, in the case of $d=3$, we have outperformed the minimal greedy algorithm with QAOA at depth $p=8$ when comparing to explicit greedy runs for small instances. However, when increasing the problem size, the minimal greedy algorithm performs strongly suboptimal. 

Our findings highlight the importance of problem structure in determining the algorithmic performance. Problems such as MaxCut, which lack an OGP regime can be effectively addressed by QAOA with a shallow number of layers. In contrast, problems with an OGP regime, like MIS, require more sophisticated quantum algorithms that break locality to achieve near-optimal solutions.

In order to boost performance by breaking locality, different forms of the mixing Hamiltonian could be considered~\cite{Tate2023}, this also allows to `warm start' from a good initial state obtained by a classical algorithm. We however also believe that hybrid quantum-classical approaches could be promising~\cite{Bravyi2019,Bae2022,Patel2022,Wagner2023,Dupont2023a,Dupont2023,Finzgar2023}. Such approaches combine quantum input, for instance $\ev{Z_i}$ and $\ev{Z_i Z_j}$ measurements on a quantum state, e.g. a QAOA state, with a classical algorithm that modifies the problem structure. In particular, when the quantum circuit is non-local and sees the whole graph such approaches may be very effective, although less reachable on the near term.
It is our hope that the tree QAOA angles will serve useful for these approaches, especially for the MIS problem, and that expected `fixed-angle' performances~\cite{Lykov2022} will also be derived for such approaches. In that way, expensive and hard parameter optimization loops can be avoided~\cite{Guerreschi2017, Zhou2018,Bittel2021}.\\

\textbf{Acknowledgements --} We thank Thomas Cope and Jalil Khatibi Moqadam for useful discussions. We also thank Thomas Cope for careful reading and comments on the manuscript. This project is supported by the Federal Ministry for Economic Affairs and Climate Action on the basis of a decision by the German Bundestag through the project Quantum-enabling Services and Tools for Industrial Applications (QuaST). QuaST aims to facilitate the access to quantum-based solutions for optimization problems.\\

\textbf{Code availability --} We provide an implementation of the recursive formulas of Appendix~\ref{ssec:add_tsym} and pre-calculated tree angles in our publicly available Python package \texttt{iqm-qaoa} \url{https://docs.meetiqm.com/iqm-qaoa/}. 

\clearpage
\appendix

\section{Calculation of the QAOA expectation values on trees} \label{sec:tree_calc}

\begin{figure} 
\centering
\includegraphics[width=0.6\textwidth]{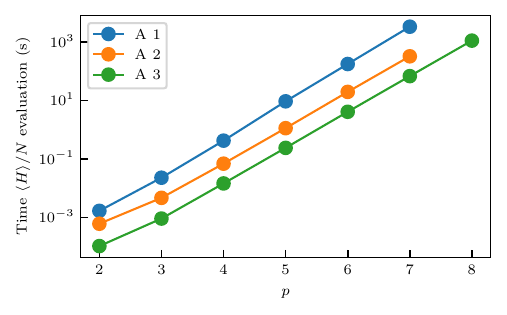}
\caption{The computational time, on a single core of a standard laptop, to evaluate Eq.~\eqref{eq:energy_density} at fixed QAOA angles using the formulas derived respectively in Sections~\ref{ssec:add_h},\ref{ssec:grow_basis_size} and~\ref{ssec:add_tsym}. The speedups obtained are prefactors, the overall scaling remains $O(2^{4p})$.} \label{fig:computation_time} 
\end{figure}

Our goal is to derive expressions that analytically describe the expectation values in Eq.~\eqref{eq:energy_density} as recursive formulas. These ideas have been pioneered in Refs.~\cite{Farhi2022,Basso2021}. In particular Ref.~\cite{Basso2021} contains very elegant expressions for the QAOA correlators in the $d \rightarrow \infty$ limit. However the finite-$d$ case received less attention in their work. The calculations presented in this work are in close correspondence to the ones presented in Ref.~\cite{Basso2021} with the following additions, each discussed in a different section: (i) We have a local field in our model~\eqref{eq:H_Ising}, and thus have no $\mathbb{Z}_2$ symmetry, resulting in less symmetric expressions, see~\ref{ssec:add_h}. (ii) We grow the basis size at every recursion step, resulting in a faster recursion, see~\ref{ssec:grow_basis_size}. (iii) We associate a symmetry label $t$ with every basis state, characterizing how `time-reversal' symmetric it is, see~\ref{ssec:add_tsym}. The corresponding computational times of evaluating Eq.~\eqref{eq:energy_density} according to the three different procedures are shown in Fig.~\ref{fig:computation_time}. In Fig.~\ref{fig:basis_iteration}, we sketch symbolically how these speedups are achieved.

\subsection{A recursive iteration with local field} \label{ssec:add_h}

In this section, our goal is to derive similar expressions as in Ref.~\cite{Basso2021} that take into account a local field. For this, we work out
\begin{multline}
\ev{Z_1 Z_2}{\bm{\gamma},\bm{\beta}}^{2\textrm{-tree}}  =  \ev{e^{i\gamma_1 H}e^{i\beta_1 B} \dots e^{i\gamma_p H}e^{i\beta_p B} Z_1 Z_2 e^{-i\beta_p B}e^{-i\gamma_p H} \dots e^{-i\beta_1 B}e^{-i\gamma_1 H}}{+},
\end{multline}
and 
\begin{equation}
\ev{Z_1}{\bm{\gamma},\bm{\beta}}^{1\textrm{-tree}}  = \ev{e^{i\gamma_1 H}e^{i\beta_1 B} \dots e^{i\gamma_p H}e^{i\beta_p B} Z_1 e^{-i\beta_p B}e^{-i\gamma_p H} \dots e^{-i\beta_1 B}e^{-i\gamma_1 H}}{+}.
\end{equation}
Here, $H$ is given by Eq.~\eqref{eq:H_Ising} and is defined on the trees shown in Fig.~\ref{fig:trees}. These trees are fully characterized by the degree $d$ and the QAOA depth $p$. As a first step, we can insert $2p+1$ resolutions of identity, labelled by superscripts $1,2,\dots,p,0,-p,\dots,-2,-1$. The basis vectors $\bm{\tilde{z}}^{[m]}$ have a dimension that is equal to the number of variables in the tree, i.e. 
\begin{equation}
\bm{\tilde{z}}^{[m]} = ( z^{[m]}_1, z^{[m]}_2,\dots, z^{[m]}_{N^{\textrm{tree}}}),\; z^{[m]}_i \in \{-1,1\} \; \textrm{with} \; i=1,\dots,N_{\textrm{tree}}
\end{equation}
For the two different tree variants, the number of variables are 
\begin{equation}
N_{2\textrm{-tree}} = 2\frac{(d-1)^{p+1}-1}{d-2}, 
\quad \quad  N_{1\textrm{-tree}} = 1+d\frac{(d-1)^{p}-1}{d-2}
\end{equation}
With this, we have that
\begin{align}
 \ev{Z_1 Z_2}{\bm{\gamma},\bm{\beta}}^{2\textrm{-tree}}& \\  =  \frac{1}{2^{N_{2\textrm{-tree}} }}  \sum_{\{\bm{\tilde{z}} \}} & \Bsl  z_1^{[0]}z_2^{[0]} \bl 
\mel{\bm{\tilde{z}}^{[1]}}{e^{i\beta_1 B}}{\bm{\tilde{z}}^{[2]}}\mel{\bm{\tilde{z}}^{[2]}}{e^{i\beta_2 B}}{\bm{\tilde{z}}^{[3]}} \dots \mel{\bm{\tilde{z}}^{[p]}}{e^{i\beta_p B}}{\bm{\tilde{z}}^{[0]}} \\
&\mel{\bm{\tilde{z}}^{[0]}}{e^{-i\beta_p B}}{\bm{\tilde{z}}^{[-p]}} \dots \mel{\bm{\tilde{z}}^{[-3]}}{e^{-i\beta_2 B}}{\bm{\tilde{z}}^{[-2]}} \mel{\bm{\tilde{z}}^{[-2]}}{e^{-i\beta_1 B}}{\bm{\tilde{z}}^{[-1]}} \br \\
& \exp(i\gamma_1 H(\bm{\tilde{z}}^{[1]}) + \dots +i\gamma_p H(\bm{\tilde{z}}^{[p]}) - i\gamma_1 H(\bm{\tilde{z}}^{[-1]}) - \dots -i\gamma_p H(\bm{\tilde{z}}^{[-p]}) ) \Bsr.
\end{align}
Here $\sum_{\{\bm{\tilde{z}}\}}$ means summing over all $2p+1$ basis sets, that each have dimension $2^{N_{2\textrm{-tree}}}$, $H(\bm{\tilde{z}}^{[m]})$ represents the energy of the bitstring. Using the same notation as in Ref.~\cite{Basso2021}, we now define the following vectors of length $(2p+1)$
\begin{equation}
\bm{\Gamma} = (\gamma_1,\dots,\gamma_p,0,-\gamma_p,\dots,-\gamma_1),
\end{equation}
and 
\begin{equation}
\bm{z}_k = ( z^{[1]}_k, \dots z^{[p]}_{k},z^{[0]}_{k},z^{[-p]}_{k},\dots z^{[-1]}_{k}),\; z^{[m]}_k \in \{-1,1\} \; \textrm{with} \; k \in V_{\textrm{tree}}.
\end{equation}
Then, we have that the mixing part of the above expression factorizes as
\begin{align}
    &\prod_{k\in V_{tree}} \bl \mel{z_k^{[1]}}{e^{i\beta_1 X_k}}{z_k^{[2]}}\mel{z_k^{[2]}}{e^{i\beta_2 X_k}}{z_k^{[3]}} \dots \mel{z_k^{[p]}}{e^{i\beta_p X_k}}{z_k^{[0]}} \\
&\mel{z_k^{[0]}}{e^{-i\beta_p X_k}}{z_k^{[-p]}} \dots \mel{z_k^{[-3]}}{e^{-i\beta_2 X_k}}{z_k^{[-2]}} \mel{z_k^{[-2]}}{e^{-i\beta_1 X_k}}{z_k^{[-1]}} \br \\
&= 2^{|V_{tree}|} \prod_{k\in V_{tree}} f(\bm{z}_k) 
\end{align}
where we defined the function $f(\bm{z}_k) \equiv f(\bm{z}_k;\beta_1,\ldots,\beta_p) $
\begin{align}
f(\bm{z}_k) = \frac{1}{2} &\mel{z_k^{[1]}}{e^{i\beta_1 X_k}}{z_k^{[2]}}\mel{z_k^{[2]}}{e^{i\beta_2 X_k}}{z_k^{[3]}} \dots \mel{z_k^{[p]}}{e^{i\beta_p X_k}}{z_k^{[0]}} \\
&\mel{z_k^{[0]}}{e^{-i\beta_p X_k}}{z_k^{[-p]}} \dots \mel{z_k^{[-3]}}{e^{-i\beta_2 X_k}}{z_k^{[-2]}} \mel{z_k^{[-2]}}{e^{-i\beta_1 X_k}}{z_k^{[-1]}}. 
\end{align}
Here, the expectation values are 
\begin{equation}
    \mel{z_k^{[m]}}{e^{i\beta_i X_k}}{z_k^{[n]}} = 
\begin{cases}
    \cos(\beta_i),& \text{if } z_k^{[m]} = z_k^{[n]}\\
    i\sin(\beta_i),              & \text{otherwise},
\end{cases}
\end{equation}
and normalization implies that $\sum_{\bm{z}_k} f(\bm{z}_k) = 1$.
With these notations, we have that
\begin{multline}\label{eq:no_tree}
\ev{Z_1 Z_2}{\bm{\gamma},\bm{\beta}}^{2\textrm{-tree}} \\ =  \sum_{\{\bm{z} \}} \Bsl z_1^{[0]}z_2^{[0]} \bl \prod_{k \in V_{2\textrm{-tree}}} f(\bm{z}_k) \br
\exp(i \frac{1}{\sqrt{d}}\sum_{kl \in E_{2\textrm{-tree}}} \bm{\Gamma} \cdot  \bm{z}_k\bm{z}_l )
\exp(i \frac{h}{\sqrt{d}}\sum_{k \in V_{2\textrm{-tree}}} \bm{\Gamma} \cdot  \bm{z}_k ) \Bsr.
\end{multline}
Here $\sum_{\{\bm{z}\}}$ should now be read as summing over all $N_{2\textrm{-tree}}$ basis sets, that each have dimension $2^{2p+1}$, $\bm{z}_k\bm{z}_l = (z^{[1]}_k z^{[1]}_l,z^{[2]}_k z^{[2]}_l,\dots)$ is the elementwise product of the two bitstrings, and $\bm{\Gamma} \cdot  \bm{z}_k $ the standard inner product between vectors.
Up to now we have not exploited the tree structure yet. This structure is however quite easy to exploit: there are only connections between generations and therefore it is natural to rearrange the summations such that we start summing from the outer children (or leaves) of the tree. Let us label an outer child by the  position index $u$, and call the parent of that child $p(u)$. 
The dependency in Eq.~\eqref{eq:no_tree} on this outer leaf can be separated as 
\begin{equation}
\sum_{\bm{z}_u} f(\bm{z}_u) \exp(i \frac{1}{\sqrt{d}}  \bm{\Gamma} \cdot  \bm{z}_u \bm{z}_{p(u)} )
\exp(i \frac{h}{\sqrt{d}} \bm{\Gamma} \cdot  \bm{z}_u). 
\end{equation}
As there are exactly $d-1$ (independent) incoming leaves in the parent node $p(u)$, we can define
\begin{equation}\label{eq:H1_first}
H_{d-1}^{(1)}(\bm{z}_{p(u)}) = \left[ \sum_{\bm{z}_u} f(\bm{z}_u) \exp(i \frac{1}{\sqrt{d}}  \bm{\Gamma} \cdot  \bm{z}_u\bm{z}_{p(u)} )
\exp(i \frac{h}{\sqrt{d}} \bm{\Gamma} \cdot  \bm{z}_u) \right]^{d-1}.
\end{equation} 
Then, we can move inwards and separate the dependencies of $\bm{z}_{p(u)}$, and so on. This will lead to the iterations 
\begin{equation}\label{eq:iteration_first}
H_{d-1}^{(m)}(\bm{a}) = \left[ \sum_{\bm{b}} f(\bm{b}) H_{d-1}^{(m-1)}(\bm{b}) \exp(i \frac{1}{\sqrt{d}}  \bm{\Gamma} \cdot  \bm{a}\bm{b})
\exp(i \frac{h}{\sqrt{d}} \bm{\Gamma} \cdot  \bm{b}) \right]^{d-1},
\end{equation}
where we have lightened the notation and now simply assume that $\bm{a}$ is a bitstring associated to a parent, and $\bm{b}$ with a child. We also have that $H_{d-1}^{(0)} \equiv 1$. After $p$ iterations, only the local fields and the edge between the two roots (labelled by 1 and 2) of the 2-tree remain, so finally we have that 
\begin{multline} \label{eq:ZZ_first}
\ev{Z_1 Z_2}{\bm{\gamma},\bm{\beta}}^{2\textrm{-tree}} \\ = \sum_{\bm{z}_1\bm{z}_2}  z_1^{[0]}z_2^{[0]} f(\bm{z}_1) f(\bm{z}_2) H_{d-1}^{(p)}(\bm{z}_1)  H_{d-1}^{(p)}(\bm{z}_2)  \exp(i \frac{1}{\sqrt{d}}  \bm{\Gamma} \cdot  \bm{z}_1\bm{z}_2) \exp(i \frac{h}{\sqrt{d}} \bm{\Gamma} \cdot  (\bm{z}_1+\bm{z}_2)).
\end{multline}
For the 1-tree, the procedure is exactly the same. However, the root (labelled by 1) has now $d$ equivalent branches. Therefore, the final iteration needs to be modified to
\begin{equation} 
H_{d}^{(p)}(\bm{a}) = \left[ \sum_{\bm{b}} f(\bm{b}) H_{d-1}^{(p-1)}(\bm{b}) \exp(i \frac{1}{\sqrt{d}}  \bm{\Gamma} \cdot  \bm{a}\bm{b})
\exp(i \frac{h}{\sqrt{d}} \bm{\Gamma} \cdot  \bm{b}) \right]^{d}.
\end{equation}
while the earlier iterations remain unchanged. 
Therefore, the local expectation value becomes
\begin{equation}
\ev{Z_1}{\bm{\gamma},\bm{\beta}}^{1\textrm{-tree}} = \sum_{\bm{z}_1}  z_1^{[0]} f(\bm{z}_1) H_{d}^{(p)}(\bm{z}_1)   \exp(i \frac{h}{\sqrt{d}} \bm{\Gamma} \cdot \bm{z}_1).
\end{equation}
The time complexity is the same for every step in the recursion in this case. Therefore the overall time complexity for the evaluation of the energy density [see Eq.~\eqref{eq:energy_density}] is $O((p+1) 2^{4p+2})$. As a difference with Ref.~\cite{Basso2021}, note that $H_{d}^{(m)}(\bm{a})$ is generally not real anymore. This is a consequence of breaking the $\mathbb{Z}_2$ symmetry when $h\neq 0$.

\begin{figure} 
\centering
\includegraphics[width=0.99\textwidth]{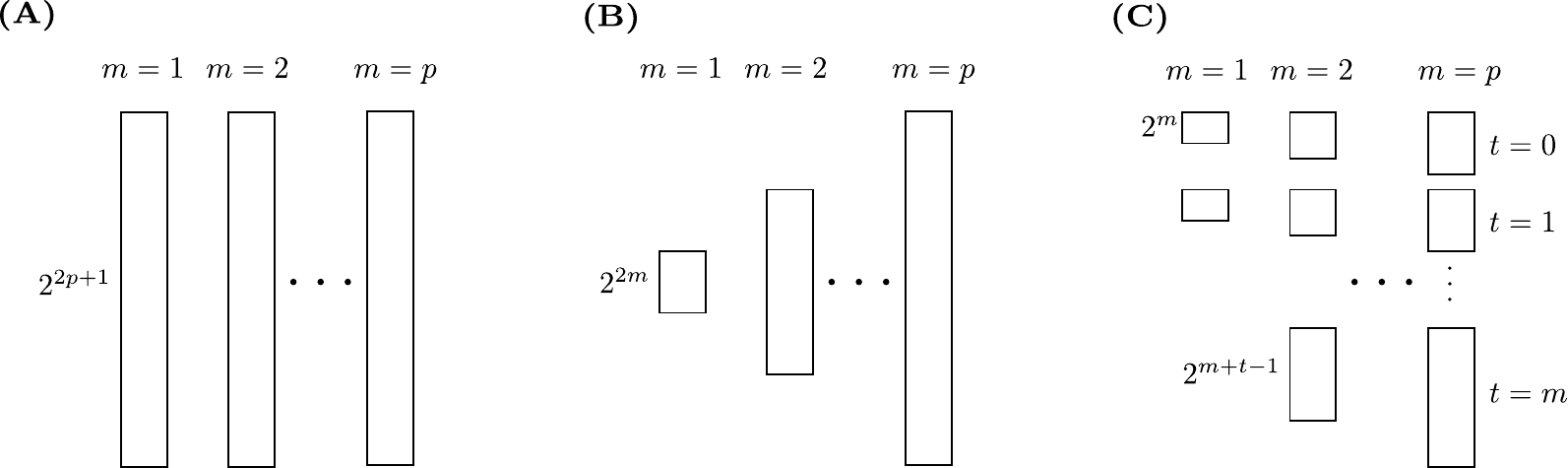}
\caption{Structure of the basis at every iteration step $m\in \{1,2,\dots,p\}$. (A) In the most naive way the basis has the same size $2^{2p+1}$ at each step. (B) For the first simplification, we grow the basis size iteratively. (C) For the second simplification, we split the basis in symmetry sectors, labelled by $t\in\{0,\dots,m\}$. } \label{fig:basis_iteration} 
\end{figure}

\subsection{First speed-up: Expanding the basis size at every iteration}\label{ssec:grow_basis_size}
In the last section, we wrote down similar expressions as in Ref.~\cite{Basso2021} with a local field included. In this section, we will discuss a first speed-up of this iterative procedure. Up to now, we did not take into account the explicit angle dependencies, that can be seen in Fig.~\ref{fig:trees}{(C)}. Taking this into account does, of course, not alter the final result, as unitary cancellations are implicit. However, taking this into account means that, for example, $H_{d-1}^{(1)}(\bm{a})$ should only depend on $\gamma_1$, $H_{d-1}^{(2)}(\bm{a})$ only on $\gamma_1,\gamma_2,\beta_1$, and so on. This is however not the case in Eq.~\eqref{eq:iteration_first}, but can be easily imposed yielding more compact iterations. 
We start by imposing the independence in Eq.~\eqref{eq:H1_first}, this gives
\begin{equation}\label{eq:H1_second}
H_{d-1}^{(1)}(\bm{a}) \rightarrow H_{d-1}^{(1)}(\bm{a}^{(1)}) =\left[ \cos(\frac{1}{\sqrt{d}}  \bm{\Gamma}^{(1)} \cdot  \bm{a}^{(1)}) \right]^{d-1}= \left[\cos(\frac{\gamma_1}{\sqrt{d}}(a_1 - a_{-1})) \right]^{d-1}
\end{equation}
where we define the vectors of length $2m$
\begin{equation}
\bm{\Gamma}^{(m)} = (\gamma_1,\dots,\gamma_m,-\gamma_m,\dots,-\gamma_1),
\end{equation}
and (to unify the notation with the QAOA angles, the subscripts now relate to the QAOA depth)
\begin{equation}\label{eq:am}
\bm{a}^{(m)} = ( a_1, \dots a_m,a_{-m},\dots a_{-1}),\; a_{i} \in \{-1,1\} .
\end{equation}
Doing the same in the general case $H_{d-1}^{(m)}(\bm{a}) \rightarrow H_{d-1}^{(m)}(\bm{a}^{(m)}) $ gives
\begin{align} \label{eq:Hd_second}
& H_{d-1}^{(m)}(\bm{a}^{(m)})  =\\ 
&\Bsl  \sum_{\bm{b}^{(m-1)}} g(\bm{b}^{(m-1)}) H_{d-1}^{(m-1)}(\bm{b}^{(m-1)})  \exp(i \frac{1}{\sqrt{d}}  \bm{\Gamma}^{(m-1)} \cdot  \bm{a}^{(m-1)}\bm{b}^{(m-1)}) 
\exp(i \frac{h}{\sqrt{d}} \bm{\Gamma}^{(m-1)} \cdot  \bm{b}^{(m-1)}) \\
&\quad\quad\frac{1}{2} \Bl  \mel{b_{m-1}}{e^{i\beta_{m-1} X}}{1} \mel{1}{e^{-i\beta_{m-1} X}}{b_{-m+1}} \exp(i\frac{\gamma_m}{\sqrt{d}}(a_m-a_{-m})) \\
&\quad\quad\quad + \mel{b_{m-1}}{e^{i\beta_{m-1} X}}{-1} \mel{-1}{e^{-i\beta_{m-1} X}}{b_{-m+1}} \exp(-i\frac{\gamma_m}{\sqrt{d}}(a_m-a_{-m}))
\Br
\Bsr^{d-1}.
\end{align}
In the last two lines, we have performed the summation over $b_{m}=b_{-m}$ explicitly. The function with reduced beta dependencies is now defined as
\begin{multline}
g(\bm{b}^{(m-1)}) \\=  \mel{b_1}{e^{i\beta_1 X}}{b_2}\dots \mel{b_{m-2}}{e^{i\beta_{m-2} X}}{b_{m-1}} 
\mel{b_{-(m-1)}}{e^{-i\beta_{m-2} X}}{b_{-(m-2)}} \dots\mel{b_{-2}}{e^{-i\beta_1 X}}{b_{-1}}.
\end{multline}
This follows from the elimination of redundant variables in $f(\bm{b})$
\begin{align} 
\sum_{\substack{b_{m+1} \dots b_{p} b_0 \\ b_{-p}  \dots b_{-(m-1)}}} f(\bm{b}) &= \frac{1}{2} \sum_{\substack{b_{m+1} \dots b_{p} b_0 \\ b_{-p}  \dots b_{-(m-1)}}} \mel{b_1}{e^{i\beta_1 X}}{b_2}\mel{b_2}{e^{i\beta_2 X}}{b_3} \dots \mel{b_p}{e^{i\beta_p X}}{b_0} \\
&\quad \quad \mel{b_0}{e^{-i\beta_p X}}{b_{-p}} \dots \mel{b_{-3}}{e^{-i\beta_2 X}}{b_{-2}} \mel{b_{-2}}{e^{-i\beta_1 X}}{b_{-1}} \\
&=  \frac{1}{2}  \mel{b_1}{e^{i\beta_1 X}}{b_2}\dots \mel{b_{m-1}}{e^{i\beta_{m-1} X}}{b_m} \braket{b_{m}}{b_{-m}} \\
&\quad \quad\mel{b_{-m}}{e^{-i\beta_{m-1} X}}{b_{-(m-1)}} \dots\mel{b_{-2}}{e^{-i\beta_1 X}}{b_{-1}} \\
&= \frac{1}{2}   g(\bm{b}^{(m-1)})   \mel{b_{m-1}}{e^{i\beta_{m-1} X}}{b_m} \mel{b_{m}}{e^{-i\beta_{m-1} X}}{b_{-(m-1)}}.
\end{align} 
Then the final result can be modified to 
\begin{align} \label{eq:ZZ_second}
& \ev{Z_1 Z_2}{\bm{\gamma},\bm{\beta}}^{2\textrm{-tree}} = \\
& \sum_{\bm{a}^{(p)} \bm{b}^{(p)} } g(\bm{a}^{(p-1)} ) g(\bm{b}^{(p-1)} ) H_{d-1}^{(p)}(\bm{a}^{(p)} )   H_{d-1}^{(p)}(\bm{b}^{(p)} )  \exp(i \frac{1}{\sqrt{d}}  \bm{\Gamma} \cdot  \bm{a}^{(p)} \bm{b}^{(p)} ) \exp(i \frac{h}{\sqrt{d}} \bm{\Gamma} \cdot  (\bm{a}^{(p)} + \bm{b}^{(p)} )) \\
&\quad\quad  \sum_{a_0 b_0} \frac{a_0b_0}{4} \mel{a_p}{e^{i\beta_{p} X}}{a_0} \mel{a_{0}}{e^{-i\beta_{p} X}}{a_{-p}} \mel{b_p}{e^{i\beta_{p} X}}{b_0} \mel{b_{0}}{e^{-i\beta_{p} X}}{b_{-p}}.
\end{align} 
The summation over $a_0$ and $b_0$ can be worked out, reducing the scaling by a prefactor to $O(2^{4p})$. The onsite term simplified in this way becomes
\begin{multline} \label{eq:Z_second}
\ev{Z_1}{\bm{\gamma},\bm{\beta}}^{1\textrm{-tree}} \\= \sum_{\bm{a}^{(p)} } g(\bm{a}^{(p-1)} ) H_{d-1}^{(p)}(\bm{a}^{(p)} )  \exp(i \frac{h}{\sqrt{d}} \bm{\Gamma} \cdot \bm{a}^{(p)} ) \sum_{a_0} \frac{a_0}{2} \mel{a_p}{e^{i\beta_{p} X}}{a_0} \mel{a_{0}}{e^{-i\beta_{p} X}}{a_{-p}}.
\end{multline}

\subsection{Second speed-up: Decomposing the basis into T-symmetric blocks} \label{ssec:add_tsym}
\begin{table}[t]
\begin{center}
\begin{tabular}{lll}
\hline\hline
$t$  & $\#$  bitstrings & bitstring\\
\hline
0&$2^m$&$\bm{a}^{(m)}_0 = (a_1,\dots,a_{m-1},a_m,a_m,a_{m-1}\dots,a_1)$  \\
1&$2^m$&$\bm{a}^{(m)}_1 =(a_1,\dots,a_{m-1},a_m,a_m,a_{m-1}\dots,-a_1)$\\
2&$2^{m+1}$&$\bm{a}^{(m)}_2 =(a_1,a_2\dots,a_{m-1},a_m,a_m,a_{m-1}\dots,-a_{2},a_{-1})$\\
$\vdots$&$\vdots$&$\vdots$  \\
$m$&$2^{2m-1}$&$\bm{a}^{(m)}_m =(a_1,\dots,a_{m-1},a_m,-a_m,a_{-(m-1)}\dots,a_{-1})$  \\
\hline\hline
\end{tabular}
\end{center}
\caption{Labelling of the basis $\{\bm{a}^{(m)}\}$, $|\{\bm{a}^{(m)}\}|=2^{2m}$, according to $t$.}
\label{tb:tsym}
\end{table}
In this section, we discuss our last speed-up that is related to a splitting of the basis in blocks that carry a symmetry label. Some properties of these been discussed in the Appendix of Ref.~\cite{Basso2021}, with as goal providing simpler iterations in the $d\rightarrow \infty$ limit, but remain unexploited for the finite $d$ iteration. 
To every bitstring of the form $\bm{a}^{(m)}$ [see Eq.~\eqref{eq:am}], we can associate a symmetry label $t\in \{0,1,\dots,m \}$, $\bm{a}^{(m)} \rightarrow \bm{a}^{(m)}_t$ that characterizes how reflection symmetric the bitstring is with respect to the middle, i.e. when $t=0$ the bitstring is completely symmetric, when $t=m$ it is not symmetric. This is summarized in Table~\ref{tb:tsym}. This label characterizes the fixed points in the iterative procedure. This is the most straightforward to see for $t=0$, where we have that $ H_{d}^{(m)}(\bm{a}^{(m)}_{0})=1$, $\forall m$.  
We can show this by induction. From Eq.~\eqref{eq:H1_second} we see that this is indeed true for $m=1$. Assuming that this is true at level $m-1$, we have for level $m$ [see Eq.~\eqref{eq:Hd_second}]
 \begin{align} 
 H_{d-1}^{(m)}(\bm{a}^{(m)}_{0}) & \\  
& = \Bsl  \sum_{\bm{b}^{(m-1)}} g(\bm{b}^{(m-1)}) \frac{1}{2}\braket{b_{m-1}}{b_{-(m-1)}} \exp(i \frac{1}{\sqrt{d}}  \bm{\Gamma}^{(m-1)} \cdot  \bm{a}^{(m-1)}_{0}\bm{b}^{(m-1)}) \\
& \quad\quad\quad \exp(i \frac{h}{\sqrt{d}} \bm{\Gamma}^{(m-1)} \cdot  \bm{b}^{(m-1)}) 
\Bsr^{d-1} \\
&=  \dots
= \Bsl  \sum_{b_1,b_{-1}}  \frac{1}{2}\braket{b_{1}}{b_{-1}} \exp(i \frac{1}{\sqrt{d}} \gamma_1 a_1 ( b_1 - b_{-1}))
\exp(i \frac{h}{\sqrt{d}} \gamma_1( b_1 - b_{-1}) ) 
\Bsr^{d-1} \\
&=  1.
\end{align}
Similarly, if $t<m$ we have that $H_{d}^{(m)}(\bm{a}^{(m)}_{t}) = H_{d}^{(t)}(\bm{a}^{(t)}_{t})$. Here $\bm{a}^{(t)}$ is obtained from $\bm{a}^{(m)}$ by pruning away the symmetric bits, i.e.
\begin{align}
&\bm{a}^{(m)}_{t} = (a_1,\dots,a_t,a_{t+1},\dots,a_m,a_m,\dots,a_{t+1},-a_t,\dots,a_{-1}) \\
&\rightarrow \bm{a}^{(t)}_{t} = (a_1,\dots,a_t,-a_t,\dots,a_{-1}).
\end{align}
The proof is similar as before, indeed, from Eq.~\eqref{eq:Hd_second} we can immediately see that a sequence of simplifications is entered when $a_m = a_{-m}$. This means that at each iteration level $m$, we `only' need to evaluate the $2^{2m-1}$ bitstrings that have $t=m$. 

We can reduce this further by a factor of two by realizing that $ H_{d-1}^{(m)}(R(\bm{a}^{(m)}_{m})) =  [H_{d-1}^{(m)}(\bm{a}^{(m)}_{m})]^{\star}$, where $R$ is the reflection operator
\begin{equation}
R:   (a_1,\dots,a_m,-a_m,\dots,a_{-1}) \rightarrow  (a_{-1},\dots,-a_t,a_t,\dots,a_{1}).
\end{equation} 

 With these ingredients, we can work out the final contraction that takes into account the symmetry sectors 
 \begin{align} \label{eq:ZZ_third}
\ev{Z_1 Z_2}{\bm{\gamma},\bm{\beta}}^{2\textrm{-tree}} &=  \sum_{t=1}^{p}  \sum_{\bm{a}^{(t)}_t \bm{b}^{(t)}_t }  G_t(a_t) G_t(b_t) g(\bm{a}^{(t-1)}_t ) g(\bm{b}^{(t-1)}_t ) H_{d-1}^{(t)}(\bm{a}^{(t)}_t )  H_{d-1}^{(t)}(\bm{b}^{(t)}_t )  \\
& \quad \quad \exp(i   \bm{\Gamma}^{(t)} \cdot \left( \frac{\bm{a}^{(t)}_t \bm{b}^{(t)}_t + h(\bm{a}^{(t)}_t + \bm{b}^{(t)}_t )}{\sqrt{d}} \right) ) \\
& \; + 2 \sum_{t_a>t_b}  \sum_{\bm{a}^{(t_a)}_{t_a} \bm{b}^{(t_a)}_{t_b} } G(a_{t_a}) \tilde{G}(b_{t_b}) g(\bm{a}^{(t_a-1)}_{t_a} ) g(\bm{b}^{(t_a-1)}_{t_b} ) H_{d-1}^{(t_a)}(\bm{a}^{(t_a)}_{t_a} )  H_{d-1}^{(t_b)}(\bm{b}^{(t_b)}_{t_b} )  \\
& \quad \quad \exp(i   \bm{\Gamma}^{(t_a)} \cdot \left( \frac{\bm{a}^{(t_a)}_{t_a} \bm{b}^{(t_a)}_{t_b} + h(\bm{a}^{(t_a)}_{t_a} + \bm{b}^{(t_a)}_{t_b} )}{\sqrt{d}} \right) ).
\end{align} 
Here, the function $G(a_t) \equiv G(a_t;\beta_t,\dots,\beta_p)$ result from the elimination of redundant symmetric variables 
\begin{align}
G(a_t) &= \sum_{a_{t+1},\dots,a_p,a_0} \frac{a_0}{2} \mel{a_t}{e^{i\beta_t X}}{a_{t+1}} \mel{a_{t+1}}{e^{-i\beta_t X}}{-a_{t}} |\mel{a_{t+1}}{e^{i\beta_{t+1}X}}{a_{t+2}}|^2 \dots |\mel{a_{p}}{e^{i\beta_{p}X}}{a_{0}}|^2 \\
&= -\frac{a_t}{2} i \sin(2\beta_t) \cos(2\beta_{t+1}) \dots \cos(2\beta_{p}).
\end{align}
Notice that if $t=p$, we have that $G(a_p)=-\frac{a_p}{2} i \sin(2\beta_p)$.
Similarly, we define 
\begin{align}
\tilde{G}(a_t) &= \sum_{a_{t+1},\dots,a_p,a_0} \frac{a_0}{2} |\mel{a_t}{e^{i\beta_t X}}{a_{t+1}}|^2 |\mel{a_{t+1}}{e^{i\beta_{t+1}X}}{a_{t+2}}|^2 \dots |\mel{a_{p}}{e^{i\beta_{p}X}}{a_{0}}|^2 \\
&= \frac{a_t}{2}  \cos(2\beta_t) \cos(2\beta_{t+1}) \dots \cos(2\beta_{p}).
\end{align}
 
For the onsite term, applying the same procedure yields
  \begin{align} \label{eq:Z_third}
\ev{Z_1}{\bm{\gamma},\bm{\beta}}^{1\textrm{-tree}} &=  \sum_{t=1}^{p}  \sum_{\bm{a}^{(t)}_t }  G_t(a_t) g(\bm{a}^{(t-1)}_t ) ) H_{d-1}^{(t)}(\bm{a}^{(t)}_t )   \exp(i \frac{h}{\sqrt{d}}  \bm{\Gamma}^{(t)} \cdot  \bm{a}^{(t)}_t ). \\
\end{align}

\subsection{The performance of tree QAOA for MaxCut relative to random sampling} \label{ssec:relative_maxcut}
In Sec.~\ref{sec:results}, in particular in Fig.~\ref{fig:fixed_h_performance}{(A)}, we reported the QAOA approximation ratios for MaxCut as defined in Sec.~\ref{sec:model}. Here, we adopted the conventional form of the MaxCut cost function~\eqref{eq:Maxcut} and defined MaxCut approximation ratios that take into account upper bounds on optimality. For this reason, we obtained non-trivial approximation ratios for random sampling, or $p=0$ QAOA, that are given by these upper bounds
\begin{equation} 
    \alpha_{MC} = \frac{c_0}{c_{ub}} = \frac{1}{2 c_{ub}}
\end{equation}
because random sampling cuts half of the edges and therefore the cut fraction is $c_0 = 1/2$. 
However, we could also be interested in the performance \textit{relative} to random sampling. For this, we can define the relative approximation ratios obtained by level-$p$ QAOA
\begin{equation} \label{eq:alpha_rel_MC}
    \alpha^{\mathrm{relative}}_{MC} = (c_p - 0.5)/(c_{ub}-0.5).
\end{equation}
Of course, this quantity can also be straightforwardly computed based on our tree calculation. The results are shown in Fig.~\ref{fig:rel_approx} and it can be seen that, similarly to local classical algorithms~\cite{Hirvonen2014}, these relative approximation ratios approach a constant in the large $d$ limit. For $p=1$, we explicitly determine the value of this constant in the next section as $\frac{1}{2\sqrt{e} P^{\star}}$ with $P^{\star}$ the Parisi constant~\ref{ssec:large_d}.

\begin{figure}[h]
    \centering
    \includegraphics[width=0.6\textwidth]{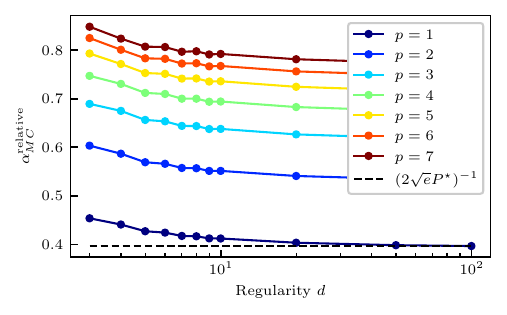}
    \caption{Relative approximation ratios achieved by QAOA that quantify the performance relative to random sampling.}
    \label{fig:rel_approx}
\end{figure}

\subsection{The performance of tree QAOA for large $d$} \label{ssec:large_d}

In this section, we investigate the performance of tree QAOA in the large $d$ limit with fixed $p=1$.
From the previous section, we can easily reduce the expectation values to $p=1$, this gives
\begin{align}
    \ev{Z_1Z_2}{{\gamma},{\beta}}^{2\textrm{-tree}} = -\frac{1}{2} \left[\sin(2\beta)\right]^2 & \left[ \cos(\frac{4h \gamma }{\sqrt{d}}) - 1 \right] \left[ \cos(\frac{2\gamma}{\sqrt{d}}) \right]^{2d-2} \\
    + & \sin(4\beta) \sin(\frac{2\gamma}{\sqrt{d}}) \cos(\frac{2h\gamma}{\sqrt{d}}) \left[ \cos(\frac{2\gamma}{\sqrt{d}}) \right]^{d-1}
\end{align}
where the two terms correspond to the two summations in Eq.~\eqref{eq:ZZ_third}, more precisely the cases where $t_a=t_b=1$ and $t_a=1$,$t_b=0$). The local term becomes
\begin{align}
    \ev{Z_1}{{\gamma},{\beta}}^{1\textrm{-tree}} &= \sin(2\beta) \sin(\frac{2h\gamma}{\sqrt{d}}) \left[ \cos(\frac{2\gamma}{\sqrt{d}}) \right]^d,
\end{align}
which is just the single case $t_a=1$ appearing in Eq.~\eqref{eq:Z_third}.

Tree QAOA with $p=1$ aims to minimize Eq.~\eqref{eq:energy_density}, i.e. the following function
\begin{align} \label{eq:p1_full}
 \frac{h}{\sqrt{d}} & \ev{Z_1}{{\gamma},{\beta}}^{1\textrm{-tree}} + \frac{\sqrt{d}}{2} \ev{Z_1 Z_2}{{\gamma},{\beta}}^{2\textrm{-tree}}  \\ = 
  & \frac{h}{\sqrt{d}} \sin(2\beta) \sin(\frac{2h\gamma}{\sqrt{d}}) \left[ \cos(\frac{2\gamma}{\sqrt{d}}) \right]^d  
     - \frac{\sqrt{d}}{4}  \left[\sin(2\beta)\right]^2 \left[ \cos(\frac{4h \gamma }{\sqrt{d}}) - 1 \right] \left[ \cos(\frac{2\gamma}{\sqrt{d}}) \right]^{2d-2} \\
&+  \frac{\sqrt{d}}{2} \sin(4\beta) \sin(\frac{2\gamma}{\sqrt{d}}) \cos(\frac{2h\gamma}{\sqrt{d}}) \left[ \cos(\frac{2\gamma}{\sqrt{d}}) \right]^{d-1} .
\end{align}
In case of MaxCut, we have $h=0$. Therefore trivially $\ev{Z_1}{{\gamma},{\beta}}^{1\textrm{-tree}} =0 $ and the correlator is given by
\begin{equation}
    \ev{Z_1 Z_2}{{\gamma},{\beta}}^{2\textrm{-tree}} = \sin(4\beta)\sin(\frac{2\gamma}{\sqrt{d}}) \left[ \cos(\frac{2\gamma}{\sqrt{d}}) \right]^{d-1}.
\end{equation}
This expression is minimized by choosing $\beta^{\star}=-\pi/8$ irrespective of $d$. The optimal gamma $\gamma^{\star}$ can be found by maximizing the remaining part which gives
\begin{equation}\label{eq:gammap1}
    \gamma^{\star} = \frac{\sqrt{d}}{2} \cos^{-1}\left(\sqrt{\frac{d-1}{d}}\right),
\end{equation}
for increasing $d$ this becomes $\gamma^{\star} \rightarrow \frac{\sqrt{d}}{2} \times \frac{1}{\sqrt{d}}=1/2 $. In this limit the expectation value becomes 
\begin{equation}\label{eq:ZZp1}
    \ev{Z_1Z_2}{{\gamma},{\beta^{\star}}}^{2\textrm{-tree}} \approx \frac{2\gamma}{\sqrt{d}}e^{-2\gamma^2}(1+O(d^{-1}))
\end{equation}
 by a Taylor expansion. Therefore, the cut fraction obtained by tree QAOA is $1/2$ in the leading order which is the same as random sampling. 
 However, as in Appendix~\ref{ssec:relative_maxcut}, we could also be interested in the performance relative to random sampling. For this we would like to compute the relative approximation ratio~\eqref{eq:alpha_rel_MC} in the large $d$ limit restricted to $p=1$ QAOA. From the above Eqs.~\eqref{eq:gammap1} and~\eqref{eq:ZZp1}, and by recalling the definition of $c_p$~\eqref{eq:cp}, we have that $c_1 - 0.5 = (2\sqrt{de})^{-1}$ for large $d$. For the denominator we can use the results of~\cite{Dembo2015} that imply that $(c_{ub}-0.5) \approx P^{\star}/\sqrt{d}$ with $P^{\star}\approx 0.7632\dots $ the Parisi constant. Therefore it follows that the relative approximation ratio converges to a constant in the large $d$ limit that can be determined as
\begin{equation}
    \alpha^{\mathrm{relative}}_{MC}(d\rightarrow \infty, p=1) = \frac{1}{2\sqrt{e} P^{\star}} \approx 0.397\dots.
\end{equation}
This is consistent with our numerical results shown in the previous section~\ref{ssec:relative_maxcut}.

In case of MIS, $h\approx d$, the single-body term $\ev{Z_1}{{\gamma},{\beta}}^{1\textrm{-tree}} $ contributes. For increasing $d$, we have that $\beta^{\star} \rightarrow -\pi/4$ and $\gamma^{\star} \rightarrow \frac{\pi}{4\sqrt{d}}$. This means that in this limit $\ev{Z_1Z_2}{{\gamma^{\star}},{\beta^{\star}}}^{2\textrm{-tree}} \rightarrow 1 $ and $\ev{Z_1}{{\gamma^{\star}},{\beta^{\star}}}^{1\textrm{-tree}} \rightarrow -1 $. Then, we have a vanishing independence ratio, and thus a vanishing MIS approximation ratio.

\clearpage

\section{Tree angles for MaxCut and MIS} \label{sec:angles_used}
We include the tree angles that were used in the finite-size simulations shown in Fig.~\ref{fig:finite_size} in the tables below. The angles below are obtained for the QAOA ansatz given in Eq.~\eqref{eq:qaoa_ansatz}, for better table alignment we however show $\bm{\beta} \rightarrow -\bm{\beta}$. For MaxCut these angles are found by taking $h=0$, and $h-2$ for MIS.

\begin{table}[H]
\begin{center}
\begin{tabular}{cl}
\hline\hline
QAOA depth $p \quad \quad$ & Tree angles for MaxCut $d=3$\\
\hline
1 &$\bm{\gamma}=$(0.5330)\\
  &$\bm{\beta}= $(0.3927)\\
2 &$\bm{\gamma}=$(0.4225, 0.7776)\\
  &$\bm{\beta}= $(0.5549, 0.2924)\\
3 &$\bm{\gamma}=$(0.3653, 0.6914, 0.8114)\\
  &$\bm{\beta}= $(0.6090, 0.4596, 0.2357)\\
4 &$\bm{\gamma}=$(0.3540, 0.6760, 0.8557, 1.0019)\\
  &$\bm{\beta}= $(0.5996, 0.4343, 0.2968, 0.1590)\\
5 &$\bm{\gamma}=$(0.3111, 0.6115, 0.7119, 0.8697, 0.9993)\\
  &$\bm{\beta}= $(0.6317, 0.5225, 0.3901, 0.2760, 0.1493)\\
6 &$\bm{\gamma}=$(0.2870, 0.5591, 0.6336, 0.7248, 0.8749, 0.9763)\\
  &$\bm{\beta}= $(0.6359, 0.5344, 0.4633, 0.3600, 0.2585, 0.1388)\\
7 &$\bm{\gamma}=$(0.2682, 0.5342, 0.5960, 0.6496, 0.7426, 0.8845, 0.9750)\\
  &$\bm{\beta}= $(0.6476, 0.5531, 0.4893, 0.4448, 0.3408, 0.2444, 0.1312)\\
8 &$\bm{\gamma}=$(0.2537, 0.5076, 0.5674, 0.6132, 0.6619, 0.7490, 0.8892, 0.9669)\\
  &$\bm{\beta}= $(0.6492, 0.5555, 0.5013, 0.4690, 0.4202, 0.3195, 0.2310, 0.1229)\\
\hline\hline
\end{tabular}
\end{center}
\end{table}
\begin{table}[H]
\begin{center}
\begin{tabular}{cl}
\hline\hline
QAOA depth $p \quad \quad$ & Tree angles for MaxCut $d=4$\\
\hline
1 &$\bm{\gamma}=$(0.5236)\\
  &$\bm{\beta}= $(0.3927)\\
2 &$\bm{\gamma}=$(0.4078, 0.7397)\\
  &$\bm{\beta}= $(0.5341, 0.2830)\\
3 &$\bm{\gamma}=$(0.3545, 0.6514, 0.7543)\\
  &$\bm{\beta}= $(0.5879, 0.4232, 0.2230)\\
4 &$\bm{\gamma}=$(0.3150, 0.5876, 0.6732, 0.7712)\\
  &$\bm{\beta}= $(0.6050, 0.4778, 0.3613, 0.1875)\\
5 &$\bm{\gamma}=$(0.2909, 0.5468, 0.6033, 0.6872, 0.7844)\\
  &$\bm{\beta}= $(0.6225, 0.5051, 0.4167, 0.3253, 0.1628)\\
6 &$\bm{\gamma}=$(0.2687, 0.5128, 0.5636, 0.6141, 0.6957, 0.7867)\\
  &$\bm{\beta}= $(0.6293, 0.5232, 0.4528, 0.3883, 0.2981, 0.1459)\\
7 &$\bm{\gamma}=$(0.2537, 0.4890, 0.5317, 0.5757, 0.6214, 0.6976, 0.7885)\\
  &$\bm{\beta}= $(0.6378, 0.5327, 0.4719, 0.4325, 0.3632, 0.2778, 0.1339)\\
8 &$\bm{\gamma}=$(0.2405, 0.4690, 0.5113, 0.5480, 0.5851, 0.6257, 0.7221, 0.8339)\\
  &$\bm{\beta}= $(0.6405, 0.5385, 0.4817, 0.4526, 0.4101, 0.3452, 0.2605, 0.1198)\\
\hline\hline
\end{tabular}
\end{center}
\end{table}
\begin{table}[ht]
\begin{center}
\begin{tabular}{cl}
\hline\hline
QAOA depth $p \quad \quad$ & Tree angles for MIS $d=3$\\
\hline
1 &$\bm{\gamma}=$(0.4299)\\
  &$\bm{\beta}= $(0.3986)\\
2 &$\bm{\gamma}=$(0.3678, 0.7957)\\
  &$\bm{\beta}= $(0.5175, 0.2642)\\
3 &$\bm{\gamma}=$(0.3260, 0.6720, 0.7582)\\
  &$\bm{\beta}= $(0.5777, 0.3680, 0.2103)\\
4 &$\bm{\gamma}=$(0.2846, 0.6045, 0.7105, 0.7419)\\
  &$\bm{\beta}= $(0.6046, 0.4686, 0.3396, 0.1846)\\
5 &$\bm{\gamma}=$(0.2722, 0.5640, 0.6536, 0.7082, 0.7908)\\
  &$\bm{\beta}= $(0.6174, 0.4776, 0.4222, 0.3088, 0.1525)\\
6 &$\bm{\gamma}=$(0.2537, 0.5324, 0.6102, 0.6370, 0.7164, 0.7872)\\
  &$\bm{\beta}= $(0.6350, 0.5082, 0.4523, 0.4136, 0.2866, 0.1533)\\
7 &$\bm{\gamma}=$(0.2316, 0.4843, 0.5687, 0.5937, 0.6238, 0.7266, 0.7577)\\
  &$\bm{\beta}= $(0.6337, 0.5026, 0.4636, 0.4304, 0.3718, 0.2725, 0.1410)\\
8 &$\bm{\gamma}=$(0.2249, 0.4734, 0.5590, 0.5833, 0.6082, 0.6697, 0.7368, 0.7882)\\
  &$\bm{\beta}= $(0.6428, 0.5181, 0.4779, 0.4519, 0.3974, 0.3589, 0.2566, 0.1203)\\
\hline\hline
\end{tabular}
\end{center}
\end{table}
\begin{table}[ht]
\begin{center}
\begin{tabular}{cl}
\hline\hline
QAOA depth $p \quad \quad$ & Tree angles for MIS $d=4$\\
\hline
1 &$\bm{\gamma}=$(0.3376)\\
  &$\bm{\beta}= $(0.4240)\\
2 &$\bm{\gamma}=$(0.3123, 0.8352)\\
  &$\bm{\beta}= $(0.5169, 0.2407)\\
3 &$\bm{\gamma}=$(0.2756, 0.6606, 0.7660)\\
  &$\bm{\beta}= $(0.5628, 0.3024, 0.1942)\\
4 &$\bm{\gamma}=$(0.2405, 0.5462, 0.6808, 0.6781)\\
  &$\bm{\beta}= $(0.5745, 0.3541, 0.2618, 0.1563)\\
5 &$\bm{\gamma}=$(0.2245, 0.5178, 0.6307, 0.6445, 0.6808)\\
  &$\bm{\beta}= $(0.6138, 0.4152, 0.3412, 0.2583, 0.1410)\\
6 &$\bm{\gamma}=$(0.2043, 0.4545, 0.5887, 0.6121, 0.6307, 0.7054)\\
  &$\bm{\beta}= $(0.6162, 0.4200, 0.3845, 0.3231, 0.2229, 0.1152)\\
7 &$\bm{\gamma}=$(0.1981, 0.4406, 0.5595, 0.5777, 0.5887, 0.6775, 0.7116)\\
  &$\bm{\beta}= $(0.6350, 0.4475, 0.4108, 0.3828, 0.3036, 0.1984, 0.1188)\\
8 &$\bm{\gamma}=$(0.1765, 0.3840, 0.5239, 0.5596, 0.5595, 0.6197, 0.6817, 0.6833)\\
  &$\bm{\beta}= $(0.6278, 0.4452, 0.4267, 0.3991, 0.3570, 0.2689, 0.1946, 0.1085)\\
\hline\hline
\end{tabular}
\end{center}
\end{table}

\newpage

\begin{figure} 
\centering
\includegraphics[width=0.99\textwidth]{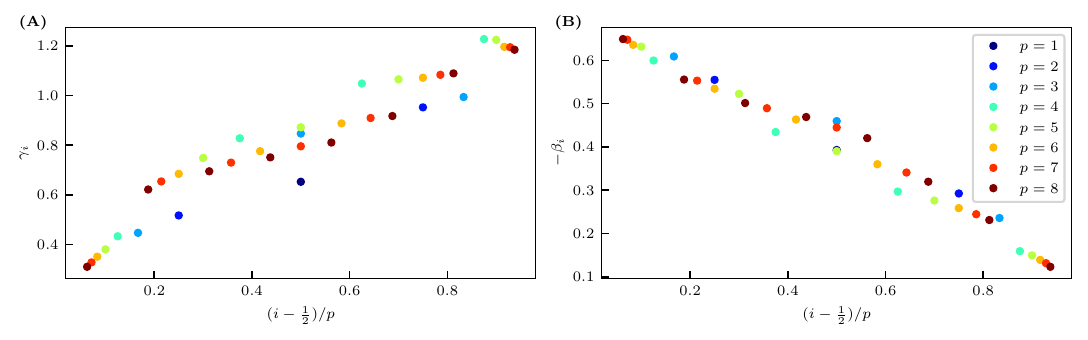}
\caption{Tree angles for MaxCut $d=3$.} \label{fig:d3_maxcut} 
\end{figure}

\begin{figure} 
\centering
\includegraphics[width=0.99\textwidth]{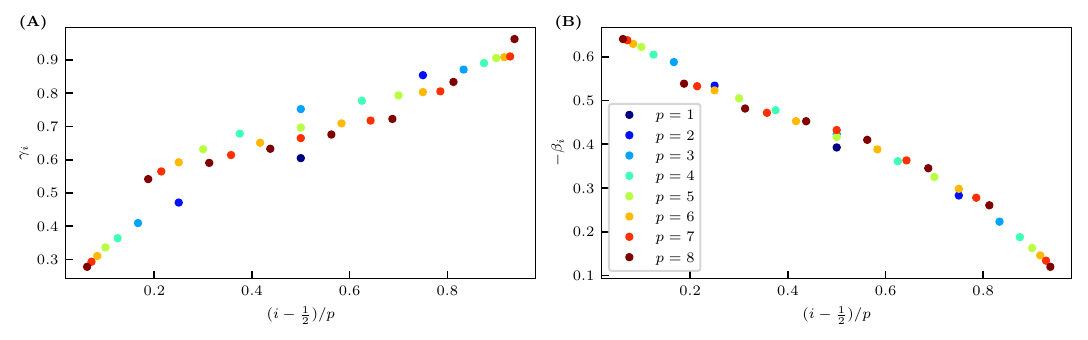}
\caption{Tree angles for MaxCut $d=4$.} \label{fig:d4_maxcut} 
\end{figure}

\begin{figure} 
\centering
\includegraphics[width=0.99\textwidth]{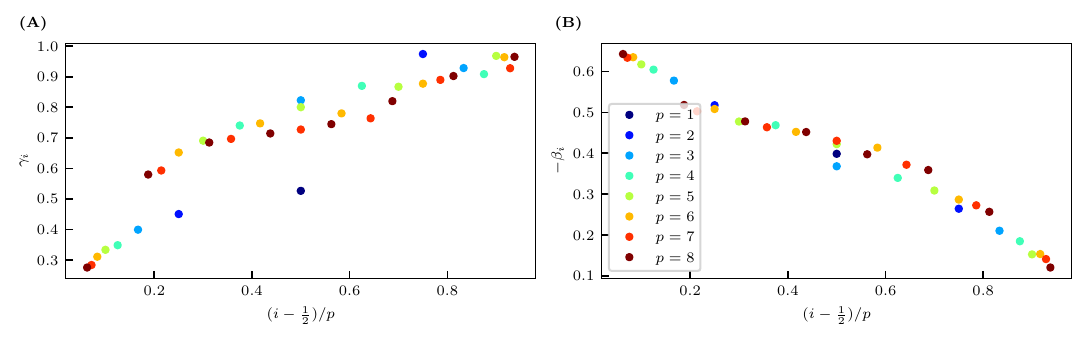}
\caption{Tree angles for MIS $d=3$.} \label{fig:d3_mis} 
\end{figure}
\begin{figure} 
\centering
\includegraphics[width=0.99\textwidth]{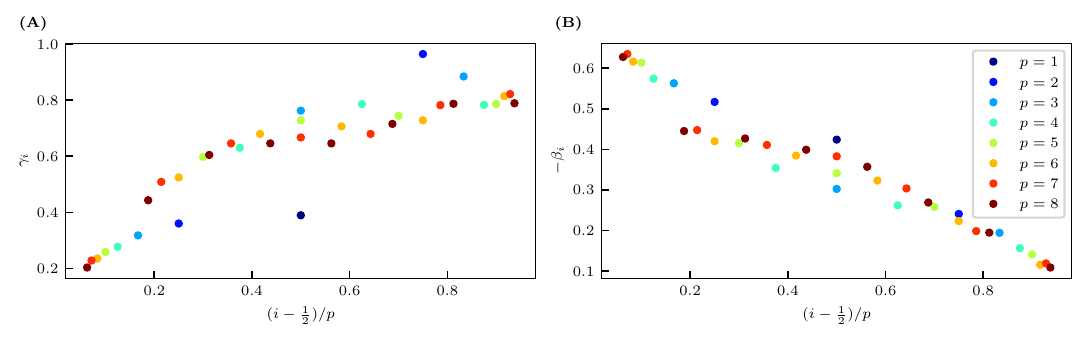}
\caption{Tree angles for MIS $d=4$.} \label{fig:d4_mis} 
\end{figure}

\clearpage
\bibliography{biblio}
\bibliographystyle{quantum}

\end{document}